\documentclass[reprint,superscriptaddress, amsmath,amssymb, aps, floatfix]{revtex4-2}

\usepackage{textcomp, gensymb}
\usepackage{amsmath}
\usepackage{xcolor}
\usepackage{graphicx}
\usepackage{dcolumn}
\usepackage{bm}
\usepackage{hyperref}
\usepackage{siunitx}
\hypersetup{colorlinks = true,linkcolor=blue,
     filecolor=blue,
     urlcolor=blue,
     citecolor = blue, pdfauthor=author}

\begin{document}
\preprint{PHYSICAL REVIEW B} 
\renewcommand{\figureautorefname}{Fig.}
\renewcommand{\tableautorefname}{Tab.}

\title{Mapping of Elastic Properties of Twisting Metamaterials onto Micropolar Continuum using Static Calculations}
\author{Brahim Lemkalli}
\email{\textcolor{black}{brahim.lemkalli@edu.umi.ac.ma}}
\affiliation{Laboratory for the Study of Advanced Materials and Applications, Department of Physics, Moulay Ismail University, B.P. 11201, Zitoune, Meknes, Morocco}

\author{Muamer Kadic}
\affiliation{Institut FEMTO-ST, UMR 6174, CNRS, Universit\'{e} de Bourgogne Franche-Comt\'{e}, 25000 Besan\c{c}on, France}

\author{Youssef El Badri}
\affiliation{Laboratory of optics, information processing, Mechanics, Energetics and Electronics, Department of Physics, Moulay Ismail University, B.P. 11201, Zitoune, Meknes, Morocco}

\author{S\'{e}bastien Guenneau}
\affiliation{UMI 2004 Abraham de Moivre-CNRS, Imperial College London, SW7 2AZ, UK}

\author{Abdenbi Bouzid}
\affiliation{Laboratory of optics, information processing, Mechanics, Energetics and Electronics, Department of Physics, Moulay Ismail University, B.P. 11201, Zitoune, Meknes, Morocco}

\author{Younes Achaoui}
\affiliation{Laboratory for the Study of Advanced Materials and Applications, Department of Physics, Moulay Ismail University, B.P. 11201, Zitoune, Meknes, Morocco}

\date{\today}

\begin{abstract}
Recent developments in the engineering of metamaterials have brought forth a myriad of mesmerizing mechanical properties that do not exist in ordinary solids. Among these, twisting metamaterials and acoustical chirality are sample-size dependent. The purpose of this work is, first, to examine the mechanical performance of a new twisting cubic metamaterial. Then, we perform a comparative investigation of its twisting behavior using the finite element method on microstructure elements computation, a phenomenological model, and we compare them to Eringen micropolar continuum. Notably, the results of the three models are in good qualitative and quantitative agreements. Finally, a systematic comparison of dispersion relations was made for the continuum and for the microstructures with different sizes in unit cells as final proof of perfect mapping. 

\end{abstract}


\maketitle

\section{Introduction}
\indent It is an unequivocal fact that technological advancements are inextricably linked to material innovations. Indeed, the advent of the concept of metamaterials has had a significant impact on the quest to explore materials with properties that are not naturally occurring nor prohibited by the conventional laws of physics \cite{Zheng2014}. Currently, metamaterials are an emerging field of research that has brought important technological and scientific advances in several disciplines. Metamaterials are defined as rationally engineered materials based in general on several compounds, which allows them to acquire unusual properties beyond those of their constituents, such as negative values for physical properties \cite{lakes2020composites, Kadic2019}. Recently, this area of research has opened up the possibility of obtaining exceptional mechanical properties previously inaccessible, thereby, coining the notion of mechanical metamaterials \cite{nicolaou2012mechanical, grima2012materials, Zheng2014, Kadic2019, Wegener2013, Fernandez-Corbaton2019}. Various types of metamaterials exhibiting unique properties have been reported \cite{ scheibner2020odd, bertoldi2017flexible, Fischer2020,  neville2016shape}. Examples of these include materials with programmable mechanical characteristics \cite{Florijn2014}, ultralight densities \cite{Schaedler2011}, and cloaking behavior \cite{Pendry2006, Buckmann2014}. Special subsets of mechanical metamaterials show features that are fundamentally beyond those of ordinary materials, for example, negative Poisson's ratio \cite{Lakes1987, Alderson2001, Mizzi2018, Duan2020}, negative compressibility \cite{Baughman1998, lakes2008negative}, and negative stiffness \cite{Janmey2006}. The tremendous advances in computer processing and the available state-of-the-art fabrication techniques have pushed the boundaries of the feasibility of implementing many complex 3D structures \cite{Yuan2021}, which has opened the prospect of exploring numerous architectural patterns that can potentially lead to new properties, such as the presence of chirality within a geometry \cite{ha2016chiral}.\newline
\indent Frenzel et \textit{al}. reported the characteristic of transforming stress into twist when the metamaterial is subjected to axial load \cite{Frenzel2017}. They have proposed an experimental investigation and numerical simulation of a 3D metamaterial formed of chiral cubic elementary cells, which has led to the emergence of a new type of mechanical metamaterials. In line with the present trend of enhancing the twisting performance of metamaterials, Chen et \textit{al}. suggested a systematic topology optimization approach for the design of tube- or beam-based metamaterials that exhibit twisting under axial stress \cite{Chen2018}. Besides, Wu et \textit{al}. explored the geometric coupling relationships between left-hand and right-hand climbing gourd tendrils and presented a tetrachiral cylindrical tube that generates a similar twisting effect under compression conditions \cite{Wu2018}. Ma et \textit{al}. reported an experimental and simulation examination of a chiral cylindrical shell twist structure in which the mechanical performance is mainly dictated by the geometrical design of the unit cell and chirality within the structure \cite{Ma2018}. Inspired by metamaterials with a negative Poisson's ratio \cite{Fu2018}, Zheng et \textit{al}. created a structure with a unidirectional twist by connecting chiral honeycomb layers with inclined rods \cite{Zheng2019}. Later, the same group observed that compression-twisting conversion are influenced by three factors: unit cell twisting, vertical superposition, and horizontal constraints \cite{Zhong2019}.\newline
\indent Xiang et \textit{al}. employed analytical and numerical simulations to establish the twisting effect of a 3D metamaterial with a non-axial square lattice \cite{Li2019}. Yan-Bin Wang et \textit{al}. made a cylindrical type structure composed of inclined rods, which has been dubbed the spring rotation structure \cite{Wang2020}. Liang Wang et \textit{al}. reported a cubic construction with double inclined rods, which was further optimized by perforating the redundant parts of the deformation of the consistent structure regularly and periodically \cite{Wang2020a}. Most recently, Frenzel et \textit{al}. have made their structure more efficient by adding arms that connect the cells to disregard the effect of horizontal constraints. Thus, they obtained a relationship between the axial deformation and the angle of twist, which is directly proportional to the length of the side of the cell \cite{Frenzel2021}. The twisting metamaterials can be employed as a novel route to perform displacement-rotation conversion, for example, transforming twisting waves into compression waves and vice versa. They have the potential to drastically improve the development of robotics, machines and even be implemented as sensors in different fields, including aerospace \cite{Frenzel2017,Zhong2019,Wang2020,Wang2020a, Frenzel2021,Wang2021,Surjadi2019}.\newline
\indent In regards to the fundamental theories involved, the traditional Cauchy theory, which is commonly applied to model 2D and 3D achiral structures, cannot describe the chiral effect, which makes it inapplicable to metamaterials with chiral structures. In general, there are two types of higher-order theories that can represent the dimensional or non-local effect caused by the material's internal microstructure: the deformation gradient theory \cite{Mindlin1968} and the microcontinuum theory \cite{lakes1982noncentrosymmetry, eringen1999theory}. The latter is further classified into micromorphic, microstretch, and micropolar (referred to as Cosserat elasticity \cite{lakes1995experimental}) theories. It is noteworthy that the micropolar continuum theory is the most extensively employed due to its mathematical convenience. The micropolar continuum possesses higher degrees of freedom since it accounts for the localized rotation of each point within the material, together with the translation factored into Cauchy elasticity. Thus, the importance of retrieving the effective parameters of this theory for mechanical metamaterials \cite{lakes2016physical, alavi2021chiral, alavi2022continualization}. In this context, Duan et \textit{al}. established the relationship between the macroscopic and microscopic mechanical characteristics of a 3D chiral cubic lattice system using micropolar continuum and homogenization theories \cite{Duan2018}. Chen et \textit{al}. exploited numerical optimization to dynamically determine the effective parameters of a micropolar continuum based on dispersion relations \cite{Chen2020}.\newline
\indent In this study, we offer an altered 3D chiral cubic structure with perforated edges that exhibits twisting effect under compression in all three dimensions, as shown in \autoref{Fig. 1}(a) and \autoref{Fig. 1}(b). Theoretical and numerical investigations by means of microstructure, phenomenological model and micropolar continuum elasticity following Eringen, using finite element method within the commercial software COMSOL Multiphysics, are performed to characterize the static and dynamic responses of the twisting behavior of this metamaterial. The microstructure model is based on solving the generalized Hooke's law (Cauchy elasticity) for the isotropic material \cite{Authier2013,Milton2002}, which generally relates the applied stress $\sigma$ to the strain $\varepsilon$; the equation is expressed as follows in Einstein summation:
\begin{equation}\label{Eq001}
\sigma_{ij}=(\frac{E\nu}{(1+\nu)(1-2\nu)} \delta_{kl}+\frac{E}{(1+\nu)} \delta_{ik}\delta_{il})\varepsilon_{kl},
\end{equation}
with $E$ is Young's modulus, $\nu$ is Poisson's ratio and $\delta_{ik}$ is the Kronecker delta symbol. The geometrical nonlinearities are often a crucial aspect of mechanical engineering, however, during this work we aim exclusively at the linear elastic regime and we apply small strains to twist the metamaterial only in the reversible regime. The geometry is re-meshed during each iteration of computation to account for the deformation.

\section{Twist metamaterial design}
\begin{center} 
\begin{figure}[!ht]
\begin{center}
\includegraphics[width=8.5cm,angle=0]{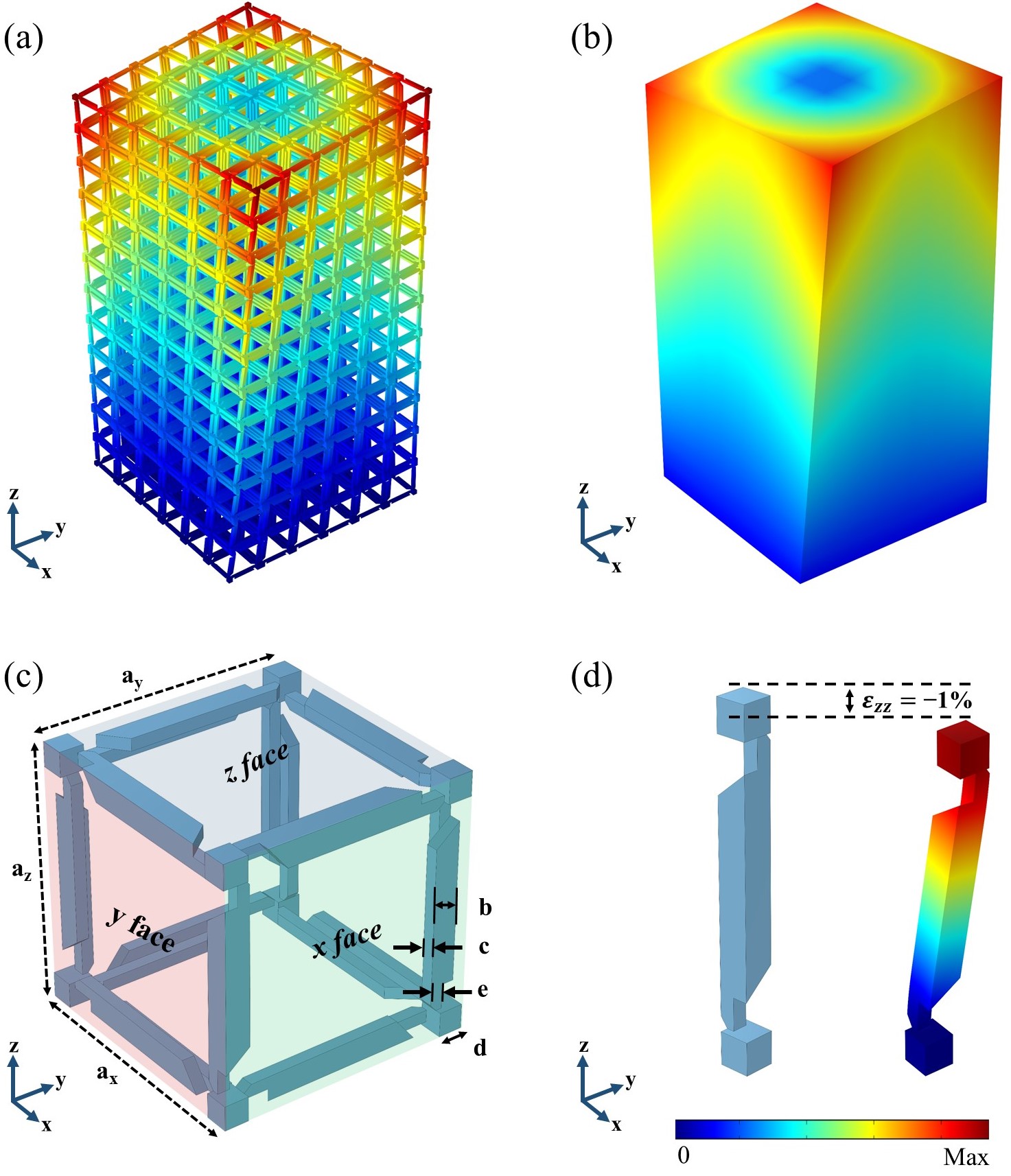}
\caption{Design of the twisting metamaterial structure. The modulus displacement vector field distributions calculated by finite element method for an axial strain of $\varepsilon_{zz}=-1\%$ along the $z$-axis: (a) metamaterials consisting of $2N^3$ unit cells ($N=6$), and (b) the effective material implemented using the micropolar continuum model with identical geometrical size to that of the microstructure. (c) The cubic unit cell with perforated edges has the following geometrical parameters $a=a_{x}=a_{y}=a_{z}=170$ \si{\mu m}, $b/a=10.29\%$, $c/a=5.88\%$, $d/a=11.76\%$, and $e/a=4.12\%$. The $8$ small-cubes in the vertices are the connectors. (d) The perforated edge deformation under $\varepsilon_{zz}=-1\%$.}
\label{Fig. 1}
\end{center}
\end{figure}
\end{center}

\indent The unit cell of the proposed metamaterial has a cubic symmetry architecture formed by arranging twelve perforated beams that permit it to twist in three directions. The structure's unit cell is connected to its adjacents via eight small cubes located in its vertices, as shown in \autoref{Fig. 1}(c). The metamaterial is composed of several unit cells that rotate in the same direction. To form a 3D metamaterial, a layer of cells is assembled in the $xy$-plane by periodically arranging $N\times N$ unit cells, $M$ layers of cells are added in the $z$-axis. The elements responsible for the twist are the four edges parallel to the axial load, as their deformation forces the superior face to rotate as shown in \autoref{Fig. 1}(d). The thickness of the perforated part of the cube's edges $e$ is the geometrical parameter that predominantly dictates the torsion angle: the narrower the perforated part, the more twist the cell will undergo. It is worth emphasizing that the connections between the unit cells within the overall structure is the determining factor that limits the twisting angle. Finally, in order to give a quantitative analysis, we have selected as constituent material the Acrylonitrile Butadiene Styrene ($ABS$), which has the following mechanical properties: $E=2.6\times10^9$ \si{Pa}, $\nu=0.4$, and $\rho=1020$ \si{kg/m^3}. This aspect has a very small influence on the study and our obtained results could be "re-scaled" to other materials. We should emphasize, however, that materials with a broad linear elastic regime are typically best suited for these applications.
\section{Microstructure computation and phenomenological model}
\begin{figure}[!ht]
\begin{center}
\includegraphics[width=7cm,angle=0]{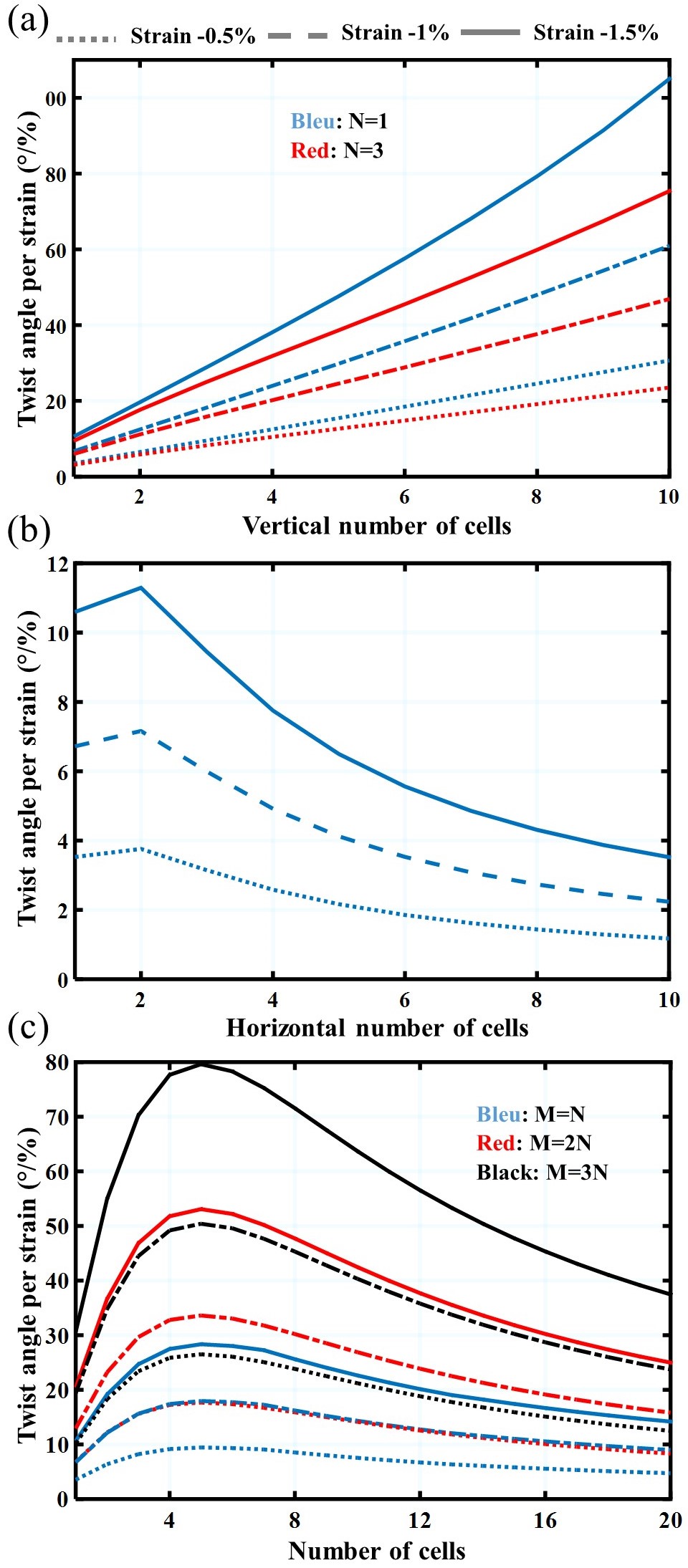}
\caption{Twist angle evolution as a function of the number of cells with three values of stain $-0.5\%$ (dotted), $-1\%$ (dashed), and $-1.5\%$ (line). (a) Vertical cells for two cases $N=1$ (blue), and $N= 3$ (red). (b) horizontal cells for $M=1$. (c) The analytical variation of the twist angle as a function of the number of cells determined from equation \ref{Eq002} for three cases $M=N$ (blue), $M=2N$ (red), and $M=3N$ (black).}
\label{Fig. 2}
\end{center}
\end{figure}

\indent In this section, we first performed simulations to delineate the variation of the twist angle as a function of the number of cells, for vertical ($M$) or horizontal ($N$) assembly of cells. The main purpose of this analysis is to explore the behavior of the metamaterial under mechanical stress and then extrapolate parameters to be implemented in the phenomenological model. We ran a series of simulations with the number of cells in the horizontal plane equal to that in the vertical plane ($M = N$), which preserves the twisting effect in all three directions in the metamaterial.\newline
\indent \autoref{Fig. 2}(a) demonstrates that the twist angle varies as a function of the number of vertical cells $M$ in a linear fashion. In addition, the twist angle increases with strain for the cases of $N=1$ and $N=3$. Thus, the twist angle for the vertical set can be written as a function of the twist angle of the unit cell as follows:
\begin{equation}\label{Eq002}
\Phi_{1,M}=M\Phi_{1,1},
\end{equation}
$\Phi_{1,1}$: twist angle of a unit cell, $\Phi_{1,M}$: twist angle for $N=1$, and $M$ cells.\newline
\indent \autoref{Fig. 2}(b) depicts the variation of the twist angle as a function of the number of horizontal cells. There is an increase of the twist angle from $\Phi_{1,1}=6.7\degree$ to $\Phi_{2,1}=7.2\degree$ and then a slower decay to a value of $\Phi_{10,1}=2.0\degree$. These results give the present metamaterial a distinct feature, namely, the increase of the twist angle in the horizontal plane from one to two horizontal cells. \newline
\indent In order to demonstrate analytically the effect of the number of cells and to describe quantitatively the twisting behavior of the metamaterial, the findings from these simulations are incorporated in the phenomenological model delineated by Frenzel et \textit{al}. that connects the strain applied to the number of unit cells based on elastic strain energy \cite{Frenzel2021}:
\begin{equation}\label{Eq003}
\Phi_{N,M}=M\frac{c^2}{N^2+c^2}\Phi_{1,1},
\end{equation}
$\Phi_{N,M}$: macrotwist angle for a metamaterial of $N$ horizontal, and $M$ vertical cells; $\Phi_{1,1}$: microrotation angle of the unit cell; $c$: number of cells characteristic constant of the metamaterial that corresponds to the maximum twist angle.\newline
\indent The analytical variation describing the effect of the number of cells on the twist angle for $M=N$, $M=2N$, and $M=3N$ configurations, which is displayed in \autoref{Fig. 2}(c). Under a $-1\%$ strain, the twist angle increases initially until it reaches its maximum, which corresponds to the number of cells $c=5$, and subsequently decreases inversely with the number of cells. To illustrate the twist angle is $15.4\degree$ for $M=N$, $33.6\degree$ for $M=2N$, and $50.4\degree$ for $M=3N$, if N=5. The analytical results demonstrate that with the increase in vertical number of cells compared to horizontal cells, the angle of twist also increases. Nevertheless, the overall twisting behavior along the three dimensions diminishes.
\indent The static response of the evolution of the twist angle as a function of the number of cells for a metamaterial composed of M=N cells is investigated and compared to the findings of the micropolar continuum discussed in the following section.
\section{Micropolar continuum}
\begin{figure*}[!ht]
\begin{center}
\includegraphics[width=14.0cm,angle=0]{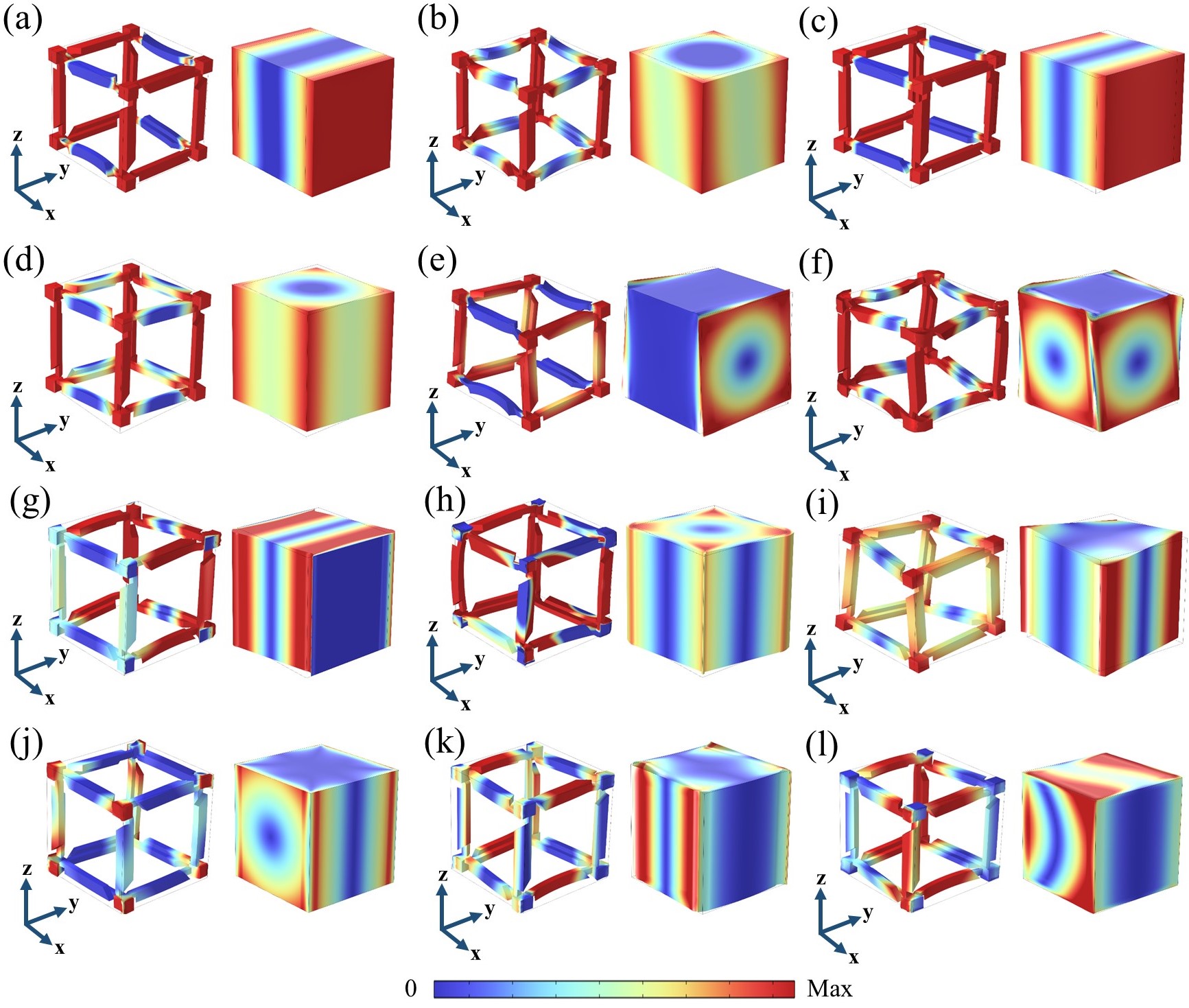}
\caption{The modulus displacement vector field distribution in the unit cell and in the continuum. (a) Test $1$: uniaxial extension. (b) Test $2$: biaxial extension. (c) Test $3$: uniaxial shear. (d) Test $4$: biaxial shear. (e) Test $5$: uniaxial twist and extension. (f) Test $6$: biaxial twist and extension. (g) Test $7$: uniaxial curvature and shear. (h) Test $8$: biaxial curvature and shear. (i) Test $9$: uniaxial twist. (j) Test $10$: biaxial twist. (k) Test $11$: uniaxial curvature. (l) Test $12$: biaxial curvature.}
\label{Fig. 3}
\end{center}
\end{figure*}
\indent In this part, we employ the theory of micropolar continuum following Eringen \cite{eringen1999theory}, in which the strain energy density for a linear elastic material is expressed by tensors of micropolar strain and curvature in Einstein summation:
\begin{equation} \label{Eq004} 
w=\frac{1}{2} \varepsilon_{ij} C_{ijkl} \varepsilon_{kl} +\varepsilon_{ij} D_{ijkl} \kappa_{kl} +\frac{1}{2} \kappa_{ij} A_{ijkl} \kappa_{kl}\,.
\end{equation} 
\indent with C, A, and D are fourth-rank normal, axial and pseudo tensors, respectively. The chiral effect is directly represented by the components of the tensor D. For each point in the medium, the strain and curvature tensors are expressed as a function of the displacement $u_{i}$, the Levi-Civita tensor $\epsilon_{jik}$ and the microrotation $\phi_{i}$, in the following form:
\begin{equation} \label{Eq005}
\varepsilon_{ij} =\frac{\partial u_{j}}{\partial x_{i}} +\epsilon_{jik} \phi_{k},
\end{equation} 
\begin{equation} \label{Eq006} 
\kappa _{ij} =\frac{\partial \phi_{i}}{\partial x_{j}}\,.
\end{equation}
\indent The strain-energy density is, by definition, a positive quantity \cite{Chen2020, Ahamdi1999}, meaning that its eigenvalues must be positive. Moreover, the stress tensors and couple stress are extracted from the strain-energy density expression by derivation in relation to the micropolar strain and the curvature tensors as follows:
\begin{equation} \label{Eq007} 
\sigma _{ij} =\frac{\partial w}{\partial \varepsilon _{ij} } =C_{ijkl} \varepsilon _{kl} +D_{ijkl} \kappa _{kl}, 
\end{equation} 
\begin{equation} \label{Eq008} 
m_{ij} =\frac{\partial w}{\partial \kappa_{ji} } =A_{jikl} \kappa_{kl} +D_{klji} \varepsilon_{kl}\,.
\end{equation}
\indent The tensors $A$, $C$, and $D$ are fourth-rank generic tensors that can be calculated using the strain energy density via the following relationships: $C_{ijkl}=\frac{\partial^2w}{\partial \epsilon_{ij}\partial \epsilon_{kl}}$, $D_{ijkl}=\frac{\partial^2w}{\partial \epsilon_{ij}\partial \kappa_{kl}}$, and $A_{ijkl}=\frac{\partial^2w}{\partial \kappa_{ij}\partial \kappa_{kl}}$. Since the structure has a chiral cubic architecture with a space group of m3m, these tensors can be reduced to four independent elements in Voigt matrix notation \cite{Chen2020,Authier2013}:
\begin{equation} \label{Eq009} 
\setlength\arraycolsep{2pt}
{\mathbf{M}=}\left(\begin{array}{ccccccccc} {{\rm M}_{11} } & {{\rm M}_{12} } & {{\rm M}_{12} } & {0} & {0} & {0} & {0} & {0} & {0} \\ {{\rm M}_{12}} & {{\rm M}_{11} } & {{\rm M}_{12} } & {0} & {0} & {0} & {0} & {0} & {0} \\ {{\rm M}_{12}} & {{\rm M}_{12}} & {{\rm M}_{11} } & {0} & {0} & {0} & {0} & {0} & {0} \\ {0} & {0} & {0} & {{\rm M}_{44} } & {0} & {0} & {{\rm M}_{47} } & {0} & {0} \\ {0} & {0} & {0} & {0} & {{\rm M}_{44} } & {0} & {0} & {{\rm M}_{47} } & {0} \\ {0} & {0} & {0} & {0} & {0} & {{\rm M}_{44} } & {0} & {0} & {{\rm M}_{47} } \\ {0} & {0} & {0} & {{\rm M}_{47}} & {0} & {0} & {{\rm M}_{44} } & {0} & {0} \\ {0} & {0} & {0} & {0} & {{\rm M}_{47}} & {0} & {0} & {{\rm M}_{44} } & {0} \\ {0} & {0} & {0} & {0} & {0} & {{\rm M}_{47}} & {0} & {0} & {{\rm M}_{44} } \end{array}\right)\,.
\end{equation} 
$M_{\alpha\beta}$ designates the general form of $C_{\alpha\beta}$, $D_{\alpha\beta}$, $A_{\alpha\beta}$, $\alpha=ij$, and $\beta=kl$.\newline
\indent To execute the micropolar continuum model, we opted for a static approach in which we employed the relationship between the elastic energy density and the effective parameters, i.e., the twelve constants of the three tensors ($C_{ijkl}$, $D_{ijkl}$, and $A_{ijkl}$). For this purpose, the twelve mechanical tests illustrated in \autoref{Fig. 3} were conducted: uniaxial and biaxial extension, as well as shear, curvature and twist, as delineated in Appendix A \cite{ Goda2015, karathanasopoulos2017designing,Duan2018,Chen2019, karathanasopoulos2020mechanics, alavi2022chiral}. The elastic energy density in the entire volume $V$ of the unit cell (\autoref{Fig. 1}(c)) is then integrated for each mechanical test. The elastic strain energy for each test is calculated as follows:
\begin{equation}\label{Eq0010}
W_{i}= \int w_{i} \,dV.\newline
\end{equation}

\begin{figure}[!htp]
\begin{center}
\includegraphics[width=8.5cm,angle=0]{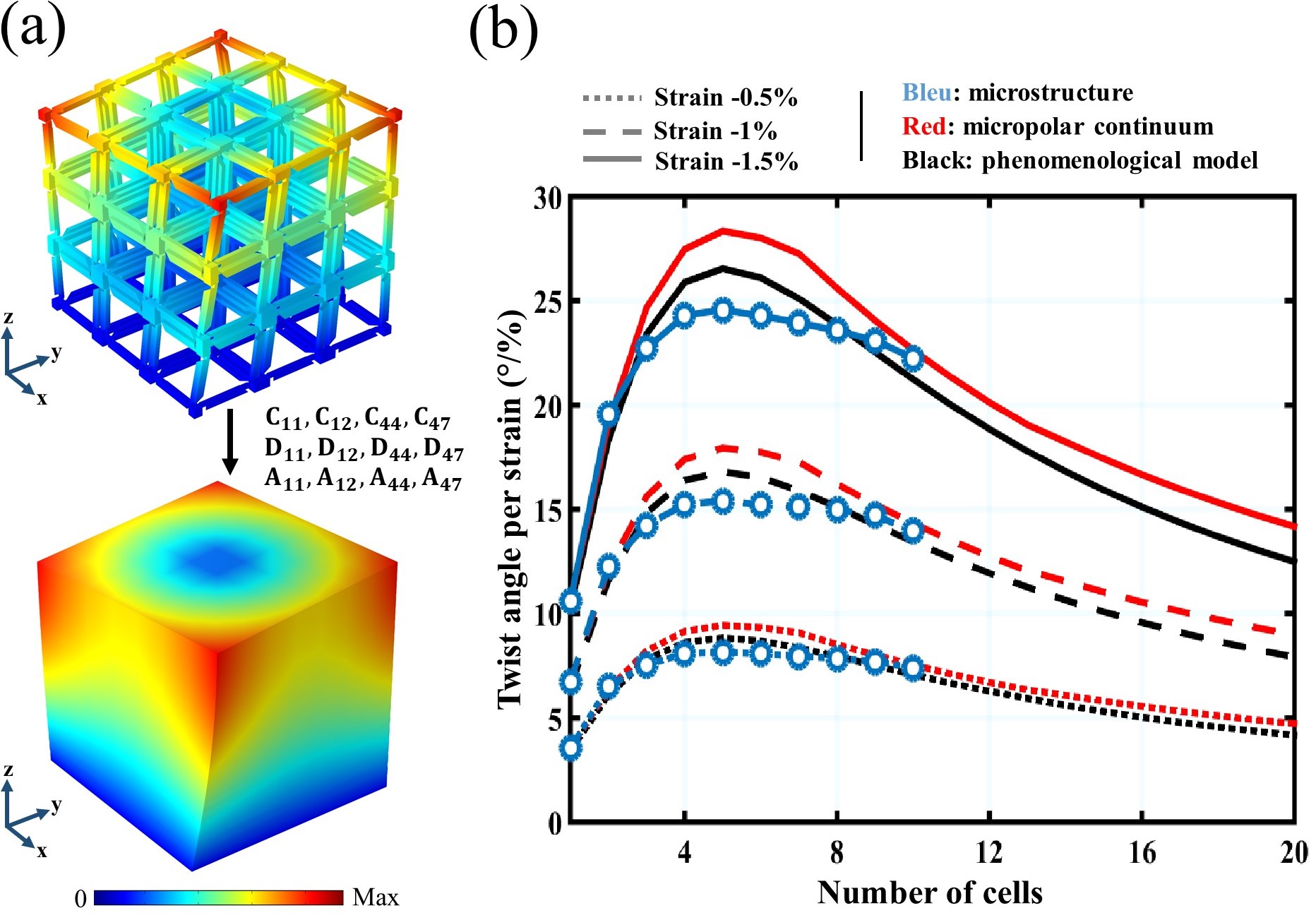}
\caption{Twist response to the axial load. (a) Total displacement field distribution for $M=N=3$ under an axial strain of $-1\%$ for the microstructure and the mapped micropolar continuum. (b) Twist angle evolution as a function of the number of cells for $M=N$ under a strain of $-0.5\%$, $-1\%$, and $-1.5\%$, described by microstructure (blue-circle), micropolar continuum (red), and the phenomenological model} (black).
\label{Fig. 4}
\end{center}
\end{figure}

\indent For each case of mechanical test, we compute the elastic energy density in the unit cell using the finite element method according to equation \ref{Eq004}, then we determine the $C_{ijkl}$, $D_{ijkl}$, and $A_{ijkl}$ constants from the relationships linking the effective parameters to the calculated elastic energies:
\begin{equation}\label{Eq011}
\left\{\begin{array}{l} {C_{11}=2W_{1}} \\ {C_{12}=W_{2}-2W_{1}} \\ {C_{44}=2W_{3}} \\ {C_{47}=W_{4}-2W_{3}} \end{array}\right.,
\end{equation}
\begin{equation}\label{Eq012}
\left\{\begin{array}{l} {D_{11}={a[\frac{6}{5}(W_{1}+W_{4}+W_{9}+W_{12}-W_{5}-W_{8})}} \\ {+\frac{4}{5}(W_{6}-W_{2}-W_{10})]}\\ {D_{12}=a[2(W_{1}+W_{9}-W_{5})+W_{6}-W_{2}-W_{10}]}\\{D_{44}=2a(W_{7}-W_{3}-W_{11})} \\ {D_{47}=a[2(W_{3}+W_{11}-W_{7})+W_{8}-W_{4}-W_{12}]} \end{array}\right.,
\end{equation}
\begin{equation}\label{Eq013}
\left\{\begin{array}{l} {A_{11}=\frac{12V}{5a} W_{9}} \\ {A_{12}=\frac{V}{a}(W_{10}-2W_{9})} \\ {A_{44}=\frac{2V}{a}W_{11}} \\ {A_{47}=\frac{V}{a}(W_{12}-2W_{11})} \end{array}\right..
\end{equation}
\indent Using the equations \ref{Eq011}, \ref{Eq012}, and \ref{Eq013}, we have determined all effective parameters of our microstructure: $C_{11}=3.845\times10^8$ \si{Pa}, $C_{12}=3.564\times10^7$ \si{Pa}, $C_{44}=7.851\times10^7$ \si{Pa}, $C_{47}=-1.064\times10^5$ \si{Pa}, $D_{11}=-0.901\times10^4$ \si{N/m}, $D_{12}=-1.478\times10^4$ \si{N/m}, $D_{44}=2.160\times10^2$ \si{N/m}, $D_{47}=-1.084\times10^3$ \si{N/m}, $A_{11}=2.01$ \si{N}, $A_{12}=1.89$ \si{N}, $A_{44}=7.552\times10^{-2}$ \si{N}, and $A_{47}=4.844\times10^{-2}$ \si{N}. \newline
\indent These parameters are used for our new Micropolar continuum in the weak form module with finite element method in COMSOL \cite{Chen2020}. This involves imposing a strain of $-1\%$ along the $z$-direction and evaluating the twist angle while concurrently fixing the displacements and micro-rotations at the bottom of the structure, for each number of cells. Say for instance, the case of a metamaterial made up of $N^3$ unitary cells with $N=3$, as shown in \autoref{Fig. 4}(a).\newline
\indent In a subsequent step, we used the micropolar continuum model, within which the statically calculated effective parameters were implemented, to determine the dispersion curves for the propagation of elastic waves along $z$-direction in the first irreducible Brillouin zone, based on the following eigenvalues equations \cite{Chen2020}:
\begin{equation}\label{Eq014}
-\omega^2 \rho u_{i} =\frac{\partial\sigma_{ji}}{\partial x_{j} },
\end{equation}
\begin{equation}\label{Eq015}
-\omega^2 \rho J \phi_{i} =\frac{\partial m_{ji}}{\partial x_{j} }+\epsilon_{ijk} \sigma_{jk},
\end{equation}
with $\rho$ is the density, $\omega$ is the angular frequency, $J$ is the micro-inertia per unit of density, we used $J=1.28 \times 10^{-8}$ \si{m^2}, and $\epsilon_{ijk}$ is the Levi-Civita tensor. The phononic dispersion was calculated by sweeping the reduced wavenumber $k_i$ in the first irreducible Brillouin zone with Floquet-Bloch periodic conditions in the $x_i$-direction as follows: $u_i(x_i+a)=u_i(x_i)e^{ik_ia}$ and $\phi_i(x_i+a)=\phi_i(x_i)e^{ik_ia}$.\\
\indent The results of the micropolar and microstructure for two cases are explored: beams consisting of $N=1$, $N=2$, $N=3$, and a bulk consisting of $N=\infty$ cells in the $xy$-plane.
\section{Discussion}
\indent We conducted a comparative analysis of the three models that have been previously described in order to investigate the static response: the microstructure computation, the phenomenological model and the micropolar continuum, which is also plotted for each effective test performed to determine the different parameters in \autoref{Fig. 3}. A comparison of their quantitative descriptions of the twisting behavior through the exploration of the evolution of the twist angle as a function of the number of cells was conducted in the case of a metamaterial with $N^3$ unit cells, as shown in \autoref{Fig. 4}(b). The blue curves represent the microstructure calculations. For a strain of $-1\%$, the twist angle increases with the number of cells, from $6.72\degree$ for $M=N=1$ to $15.4\degree$ for $M=N=5$ and then gradually decreases until stabilizing in the range of $N=5$ to $N=8$, with a constant twist angle in the vicinity of $15\degree$. It follows that the twist angle varies with the deformation: in the case of $N=5$ and for strains of $-0.5\%$, $-1\%$, and $-1.5\%$, the twist angle equals $8.83\degree$, $16.80\degree$, and $26.54\degree$, respectively. The second study corresponds to the phenomenological model, which is shown by the black curves. The twist angle increases to a maximum for $N=5$ and then gradually decays as the number of cells increases. Regarding the results of the micropolar continuum model that are indicated by the red curves, it is observed that the twist angle increases and then drops by the same rate as the phenomenological model and the microstructure computations. Thus, these results reveal that the effective parameters reported accurately reflect the static twisting behavior of the metamaterial, particularly, for small strains, which fall within the linear elastic regime. On the other hand, the difference between the three models became much more substantial at higher strains, as a result of the high degree of nonlinearity associated with the large geometrical deformation, which impacted the fitting of the three approaches we used.

\begin{figure}[!ht]
\begin{center}
\includegraphics[width=8.0cm,angle=0]{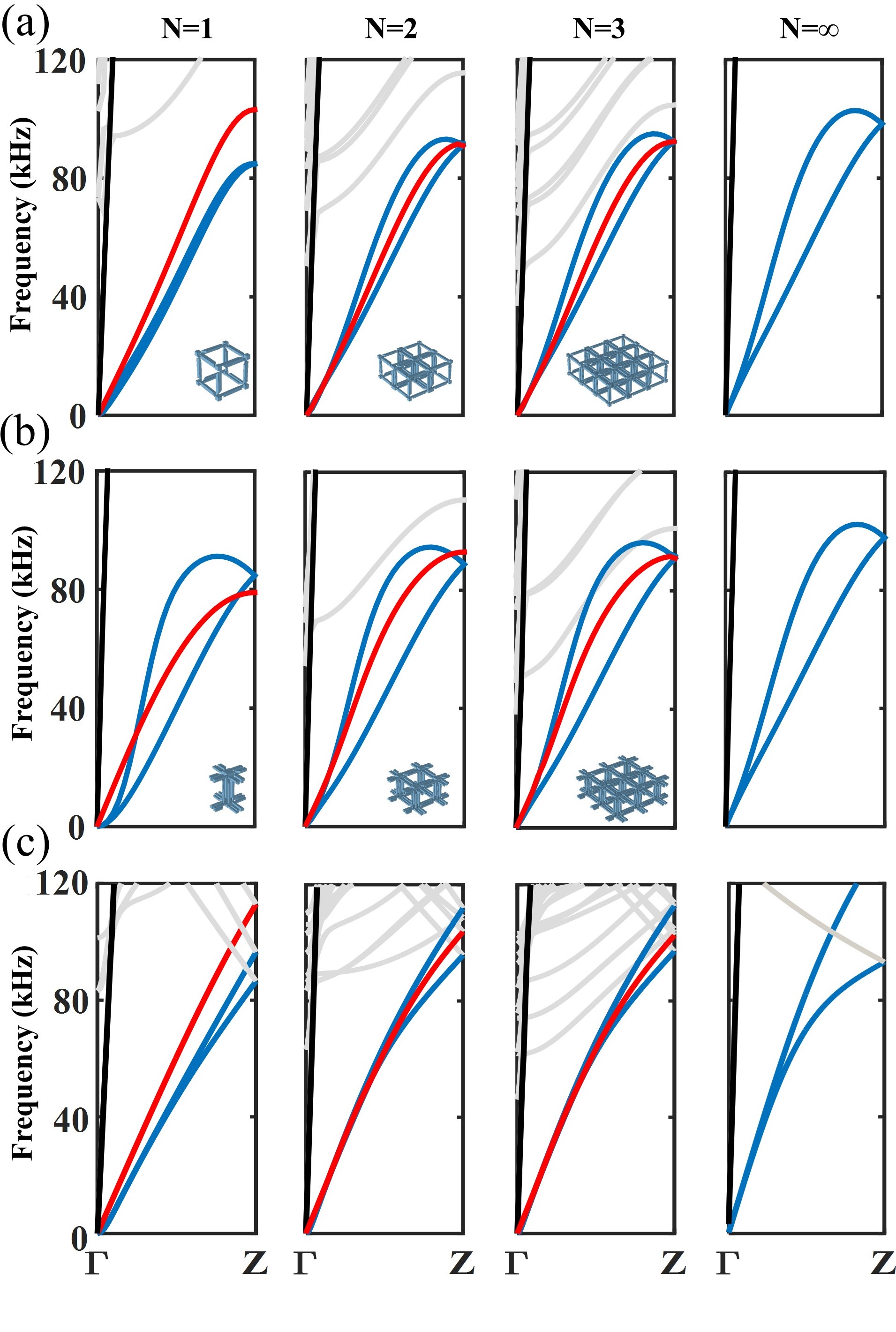}
\caption{The phononic dispersion diagrams for the first irreducible Brillouin zone along z direction with the variation of the number of cells for four cases $N=1$, $N=2$, $N=3$, and $N=\infty$. (a) Microstructure of the standard unit cell (inset in the first row). (b) Microstructure of the unit cell shaped as I (inset in the second row). (c) Micropolar continuum elasticity (third row).}
\label{Fig. 5}
\end{center}
\end{figure}
\indent Before proceeding to the comparison of the dynamic response between the two models, we opted to also introduce a customized unit cell shaped as I, and study its dispersion diagram through the microstructure model as a reference, along with the conventional unit cell previously introduced in \autoref{Fig. 1}(c). The results of this study are illustrated in \autoref{Fig. 5}, which shows the dispersion diagrams for $N=1$, $N=2$, $N=3$, and $N=\infty$ for the conventional cell (\autoref{Fig. 5}(a)) and the new customized unit cell (\autoref{Fig. 5}(b)). At a certain angular frequency, the first two modes in blue, which correspond to the bending and transverse modes, propagate with different velocities, a phenomenon known as lifting of degeneracy \cite{Chen2020}. As a result, there is an obvious degeneracy lifting rise as the number of cells increases for the standard cell. However, for the customized configuration, the degeneracy lifting is more pronounced for a beam of $N=1$ and is more resilient to the increase in the number of cells. The twist mode in red appears for beams with a finite number of cells, then disappears for $N=\infty$ for the two configurations.\newline
\indent For the sake of further confirming the accuracy of the constituted micropolar continuum model based on static mapping of effective parameters, we explore the band structures in the case of a beams ($N=1,2,3,\infty$) through the microstructure and micropolar models, as shown in \autoref{Fig. 5}(c). The first two curves highlighted in blue correspond to the transverse and bending modes, which have undergone degeneracy lifting in the fining of both microstructure and micropolar models. Thus, the weak form micropolar model successfully described the so-called acoustic activity, analogous to optical activity, which is the capacity of the twisting metamaterial to rotate linearly transversely polarized during wave propagation \cite{Chen2020}. Furthermore, the red curve, which portrays the twist mode created by the twisting motion of the beam and the other gray-highlighted curves above it, are all in good agreement. For an infinite number of cells in the $xy$-plane ($N=\infty$), the metamaterial becomes bulkier, forcing the twist mode to disappear and the two transverse and bending modes become more separated. In other words, the acoustic activity becomes more pronounced. \newline
\indent The dispersion curves for the micropolar calculations do indeed match those of the standard cell evaluated using the microstructure. The transverse, bending, twist and longitudinal modes are in good agreement. However, for the customized cell, the modes do not fit together, due to the presence of the degeneracy lifting in $N=1$. These observations show that the micropolar medium describes the microstructure to which we carried out the mechanical tests. These findings reveal unequivocally that the effective parameters determined by static mechanical tests accurately describe the dynamic behavior of the microstructure, as demonstrated by the excellent concordance between the microstructure and micropolar continuum results.
\section{Conclusion}
\indent In conclusion, we have provided various numerical results for the development of a new 3D twist metamaterial with a chiral cubic architecture. Numerical calculations based on the finite element method, analytical modeling and micropolar continuum established by Eringen were employed to model the twisting behavior of the metamaterial. The outcomes of the three approaches are in good agreement. The structure exhibits substantially superior twisting behavior, compared to previously reported twist metamaterials in the literature. Furthermore, the suggested configuration addresses one of the key challenges with twist metamaterials: the decrease in twist angle when assembling multiple cells in the horizontal plane. Likewise, in terms of longitudinal grouping, the number of cells assembled longitudinally is proportional to the twist angle, meaning that the structure retains the same properties.\\
\indent Under $-1\%$ strain, the suggested structure has a maximum twist angle of $15.4\degree$ (for $M=N$) in any of the three directions $x$, $y$, or $z$. The phenomenological model projected that under $-1\%$ deformation and for $N=5$ the twist angle goes from $15.4\degree$ for $M=N$ to $33.6\degree$ for $M=2N$, and $50.4\degree$ for $M=3N$. Thus, our metamaterial presents a new candidate for a broad range of mechanical applications that involve converting axial mechanical stress into twist and vice versa.\newline
\indent The effective parameters of the micropolar continuum have been evaluated statically through different mechanical tests: stress, uniaxial and triaxial compressions as well as bendings and twists. We have conclusively elucidated that the determination of the micropolar continuum model's effective parameters by means of static mechanical testing is a viable approach to adequately describe both the static and dynamic aspects of the twisting behavior of the metamaterial.
\section*{Acknowledgements}
M.K. is grateful for the support of the EIPHI Graduate School (Contract No. ANR-17-EURE-0002) and by the French Investissements d’Avenir program, project ISITEBFC (Contract No. ANR-15-IDEX-03).

\appendix

\section{}
\indent \textbf{Test 1} - Uniaxial extension corresponds to the energy density $W_{1}$: we apply a uniform deformation $\varepsilon_{11}=1$ on the $x$-face and a zero stress on the other $y$- and $z$-faces of the unit cell, the corresponding boundary conditions are then written as follows:\newline
\noindent - face $x$: \quad $u_{x}=x$,\newline
\noindent - faces $y$ and $z$: \quad $u_{y}=0$ and $u_{z}=0$, respectively.\\
\indent \textbf{Test 2} - Biaxial extension corresponds to the energy density $W_{2}$: we apply a uniform deformation $\varepsilon_{11}=1$ on the $x$-face and $\epsilon_{22}=1$ on the $y$-face and a zero stress on the $z$-face of the unit cell, the corresponding boundary conditions are then written as follows:\newline
\noindent - face $x$: \quad $u_{x}=x$,\newline
\noindent - face $y$: \quad $u_{y}=y$,\newline
\noindent - face $z$: \quad $u_{z}=0$.\\
\indent \textbf{Test 3} - Uniaxial shear corresponds to the energy density $W_{3}$: we apply a uniform deformation $\varepsilon_{12}=1$ on the $y$-face and a zero stress on the other $x$- and $z$-faces of the unit cell, the corresponding boundary conditions are then written as follows:\newline
\noindent - faces $x$ and $z$: \quad $u_{x}=0$ and $u_{z}=0$, respectively.\newline
\noindent - face $y$: \quad $u_{y}=x$.\\
\indent \textbf{Test 4} - Biaxial shear corresponds to the energy density $W_{4}$: we apply a uniform deformation $\varepsilon_{12}=1$ on the $y$-face and $\varepsilon_{21}=1$ on the $x$-face and a zero stress on the $z$-face of the unit cell, the corresponding limit conditions are then written as follows:\newline
\noindent - face $x$: \quad $u_{x}=y$,\newline
\noindent - face $y$: \quad $u_{y}=x$,\newline
\noindent - face $z$: \quad $u_{z}=0$.\\
\indent \textbf{Test 5} - Uniaxial twist and extension corresponds to the energy density $W_{5}$: we apply a uniform deformation $\varepsilon_{11}=1$ and $\kappa_{11}=1$ on the $x$-face and a zero stress on the other $y$- and $z$-faces of the unit cell, the corresponding limit conditions are then written as follows:\newline
\noindent - face $x$: \quad $u_{x}=x$, $u_{y}=-xz$, and $u_{z}=xy$, \newline
\noindent - faces $y$ and $z$: \quad $u_{y}=0$ and $u_{z}=0$, respectively.\\
\indent \textbf{Test 6} - Biaxial twist and extension corresponds to the energy density $W_{6}$: we apply a uniform deformation $\varepsilon_{11}=1$ et $\kappa_{11}=1$ on the $x$-face and $\varepsilon_{22}=1$ and $\kappa_{22}=1$ on the $y$-face and a zero stress on the $z$-face of the unit cell, the corresponding limit conditions are then written as follows:\newline
\noindent - face $x$: \quad $u_{x}=x$, $u_{y}=-xz$, and $u_{z}=xy$, \newline
\noindent - face $y$: \quad $u_{x}=yz$, $u_{y}=y$, and $u_{z}=-xy$, \newline
\noindent - face $z$: \quad $u_{z}=0$.\\
\indent \textbf{Test 7} - Uniaxial curvature and shear corresponds to the energy density $W_{7}$: we apply a uniform deformation $\varepsilon_{12}=1$ on the $y$-face and $\kappa_{12}=1$ on both $x$ and $y$-faces and a zero stress on the $z$-face of the unit cell, the corresponding limit conditions are then written as follows:\newline
\noindent - face $x$: \quad $u_{x}=-\frac{y^2}{2}$, $u_{y}=0$, and $u_{z}=0$, \newline
\noindent - face $y$: \quad $u_{x}=0$, $u_{y}=x+xy$, and $u_{z}=0$, \newline
\noindent - face $z$: \quad $u_{z}=0$.\\
\indent \textbf{Test 8} - Biaxial curvature and shear corresponds to the energy density $W_{8}$: we apply a uniform strain $\varepsilon_{12}=1$ on the $y$-face and $\varepsilon_{21}=1$ on the $x$-face and $\kappa_{12}=1$ et $\kappa_{21}=1$ on the three $x$-faces, $y$ and $z$, the corresponding boundary conditions are then written as follows:\newline
\noindent - face $x$: \quad $u_{x}=y-\frac{y^2}{2}$, $u_{y}=0$, and $u_{z}=0$, \newline
\noindent - face $y$: \quad $u_{x}=0$, $u_{y}=x+xy-\frac{z^2}{2}$, and $u_{z}=0$, \newline
\noindent - face $z$: \quad $u_{z}=yz$.\\
\indent \textbf{Test 9} - Uniaxial twist corresponds to the energy density $W_{9}$: we apply a uniform strain $\kappa_{11}=1$ on the $x$-face and a zero stress on the other $y$- and $z$-faces of the unit cell, the corresponding boundary conditions are then written as follows:\newline
\noindent - face $x$: \quad $u_{x}=0$, $u_{y}=-xy$, and $u_{z}=xy$, \newline
\noindent - faces $y$ and $z$: \quad $u_{y}=0$ and $u_{z}=0$, respectively.\\
\indent \textbf{Test 10} - Biaxial twist corresponds to the energy density $W_{10}$: we apply a uniform strain $\kappa_{11}=1$ on the $x$-face and $\kappa_{22}=1$ on the $y$ face and a zero stress on the $z$-face of the unit cell, the corresponding limit conditions are then written as follows:\newline
\noindent - face $x$: \quad $u_{x}=0$, $u_{y}=-xy$, and $u_{z}=xy$, \newline
\noindent - face $y$: \quad $u_{x}=yz$, $u_{y}=0$, and $u_{z}=-xy$, \newline
\noindent - face $z$: \quad $u_{z}=0$.\\
\indent \textbf{Test 11} - Uniaxial curvature corresponds to the energy density $W_{11}$: we apply a uniform strain $\kappa_{12}=1$ on both sides $x$ and $y$ and a zero stress on the $z$-face of the unit cell, the corresponding boundary conditions are then written as follows:\newline
\noindent - face $x$: \quad $u_{x}=-\frac{y^2}{2}$, $u_{y}=0$, and $u_{z}=0$, \newline
\noindent - face $y$: \quad $u_{x}=0$, $u_{y}=xy$, and $u_{z}=0$, \newline
\noindent - face $z$: \quad $u_{z}=0$.\\
\indent \textbf{Test 12} - Biaxial curvature corresponds to the energy density $W_{12}$: we apply a uniform strain $\kappa_{12}=1$ and$\kappa_{21}=1$ on the three $x$-, $y$- and $z$-faces, the corresponding boundary conditions are then written as follows:\newline
\noindent - face $x$: \quad $u_{x}=-\frac{y^2}{2}$, $u_{y}=0$, and $u_{z}=0$, \newline
\noindent - face $y$: \quad $u_{x}=0$, $u_{y}=xy--\frac{z^2}{2}$, and $u_{z}=0$, \newline
\noindent - face $z$: \quad $u_{z}=yz$.\newline
\bibliography{mybibfile}

\providecommand{\noopsort}[1]{}\providecommand{\singleletter}[1]{#1}%
\begin{thebibliography}{54}%
\makeatletter
\providecommand \@ifxundefined [1]{%
 \@ifx{#1\undefined}
}%
\providecommand \@ifnum [1]{%
 \ifnum #1\expandafter \@firstoftwo
 \else \expandafter \@secondoftwo
 \fi
}%
\providecommand \@ifx [1]{%
 \ifx #1\expandafter \@firstoftwo
 \else \expandafter \@secondoftwo
 \fi
}%
\providecommand \natexlab [1]{#1}%
\providecommand \enquote  [1]{``#1''}%
\providecommand \bibnamefont  [1]{#1}%
\providecommand \bibfnamefont [1]{#1}%
\providecommand \citenamefont [1]{#1}%
\providecommand \href@noop [0]{\@secondoftwo}%
\providecommand \href [0]{\begingroup \@sanitize@url \@href}%
\providecommand \@href[1]{\@@startlink{#1}\@@href}%
\providecommand \@@href[1]{\endgroup#1\@@endlink}%
\providecommand \@sanitize@url [0]{\catcode `\\12\catcode `\$12\catcode
  `\&12\catcode `\#12\catcode `\^12\catcode `\_12\catcode `\%12\relax}%
\providecommand \@@startlink[1]{}%
\providecommand \@@endlink[0]{}%
\providecommand \url  [0]{\begingroup\@sanitize@url \@url }%
\providecommand \@url [1]{\endgroup\@href {#1}{\urlprefix }}%
\providecommand \urlprefix  [0]{URL }%
\providecommand \Eprint [0]{\href }%
\providecommand \doibase [0]{https://doi.org/}%
\providecommand \selectlanguage [0]{\@gobble}%
\providecommand \bibinfo  [0]{\@secondoftwo}%
\providecommand \bibfield  [0]{\@secondoftwo}%
\providecommand \translation [1]{[#1]}%
\providecommand \BibitemOpen [0]{}%
\providecommand \bibitemStop [0]{}%
\providecommand \bibitemNoStop [0]{.\EOS\space}%
\providecommand \EOS [0]{\spacefactor3000\relax}%
\providecommand \BibitemShut  [1]{\csname bibitem#1\endcsname}%
\let\auto@bib@innerbib\@empty
\bibitem [{\citenamefont {Zheng}\ \emph {et~al.}(2014)\citenamefont {Zheng},
  \citenamefont {Lee}, \citenamefont {Weisgraber}, \citenamefont {Shusteff},
  \citenamefont {Deotte}, \citenamefont {Duoss}, \citenamefont {Kuntz},
  \citenamefont {Biener}, \citenamefont {Ge}, \citenamefont {Jackson},
  \citenamefont {Kucheyev}, \citenamefont {Fang},\ and\ \citenamefont
  {Spadaccini}}]{Zheng2014}%
  \BibitemOpen
  \bibfield  {author} {\bibinfo {author} {\bibfnamefont {C.}~\bibnamefont
  {Zheng}}, \bibinfo {author} {\bibfnamefont {H.}~\bibnamefont {Lee}}, \bibinfo
  {author} {\bibfnamefont {T.~H.}\ \bibnamefont {Weisgraber}}, \bibinfo
  {author} {\bibfnamefont {M.}~\bibnamefont {Shusteff}}, \bibinfo {author}
  {\bibfnamefont {J.}~\bibnamefont {Deotte}}, \bibinfo {author} {\bibfnamefont
  {E.~B.}\ \bibnamefont {Duoss}}, \bibinfo {author} {\bibfnamefont {J.~D.}\
  \bibnamefont {Kuntz}}, \bibinfo {author} {\bibfnamefont {M.~M.}\ \bibnamefont
  {Biener}}, \bibinfo {author} {\bibfnamefont {Q.}~\bibnamefont {Ge}}, \bibinfo
  {author} {\bibfnamefont {J.~A.}\ \bibnamefont {Jackson}}, \bibinfo {author}
  {\bibfnamefont {S.~O.}\ \bibnamefont {Kucheyev}}, \bibinfo {author}
  {\bibfnamefont {N.~X.}\ \bibnamefont {Fang}},\ and\ \bibinfo {author}
  {\bibfnamefont {C.~M.~.}\ \bibnamefont {Spadaccini}},\ }\bibfield  {title}
  {\bibinfo {title} {{Ultralight, ultrastiff mechanical metamaterials}},\
  }\href {https://doi.org/10.1126/science.1252291} {\bibfield  {journal}
  {\bibinfo  {journal} {Science}\ }\textbf {\bibinfo {volume} {344}},\ \bibinfo
  {pages} {6190} (\bibinfo {year} {2014})}\BibitemShut {NoStop}%
\bibitem [{\citenamefont {Lakes}(2020)}]{lakes2020composites}%
  \BibitemOpen
  \bibfield  {author} {\bibinfo {author} {\bibfnamefont {R.}~\bibnamefont
  {Lakes}},\ }\href {https://doi.org/10.1142/11715} {\emph {\bibinfo {title}
  {Composites and metamaterials}}}\ (\bibinfo  {publisher} {World Scientific},\
  \bibinfo {year} {2020})\BibitemShut {NoStop}%
\bibitem [{\citenamefont {Kadic}\ \emph {et~al.}(2019)\citenamefont {Kadic},
  \citenamefont {Milton}, \citenamefont {van Hecke},\ and\ \citenamefont
  {Wegener}}]{Kadic2019}%
  \BibitemOpen
  \bibfield  {author} {\bibinfo {author} {\bibfnamefont {M.}~\bibnamefont
  {Kadic}}, \bibinfo {author} {\bibfnamefont {G.~W.}\ \bibnamefont {Milton}},
  \bibinfo {author} {\bibfnamefont {M.}~\bibnamefont {van Hecke}},\ and\
  \bibinfo {author} {\bibfnamefont {M.}~\bibnamefont {Wegener}},\ }\bibfield
  {title} {\bibinfo {title} {{3D metamaterials}},\ }\href
  {https://doi.org/10.1038/s42254-018-0018-y} {\bibfield  {journal} {\bibinfo
  {journal} {Nature Reviews Physics 2019 1:3}\ }\textbf {\bibinfo {volume}
  {1}},\ \bibinfo {pages} {198} (\bibinfo {year} {2019})}\BibitemShut {NoStop}%
\bibitem [{\citenamefont {{Nicolaou, Zachary G and Motter, Adilson
  E}}(2012)}]{nicolaou2012mechanical}%
  \BibitemOpen
  \bibfield  {author} {\bibinfo {author} {\bibnamefont {{Nicolaou, Zachary G
  and Motter, Adilson E}}},\ }\bibfield  {title} {\bibinfo {title} {{Mechanical
  metamaterials with negative compressibility transitions}},\ }\href
  {https://doi.org/10.1038/nmat3331} {\bibfield  {journal} {\bibinfo  {journal}
  {Nature materials}\ }\textbf {\bibinfo {volume} {11}},\ \bibinfo {pages}
  {608} (\bibinfo {year} {2012})}\BibitemShut {NoStop}%
\bibitem [{\citenamefont {{Grima, Joseph N and Caruana-Gauci,
  Roberto}}(2012)}]{grima2012materials}%
  \BibitemOpen
  \bibfield  {author} {\bibinfo {author} {\bibnamefont {{Grima, Joseph N and
  Caruana-Gauci, Roberto}}},\ }\bibfield  {title} {\bibinfo {title} {{Materials
  that push back}},\ }\href {https://doi.org/10.1038/nmat3364} {\bibfield
  {journal} {\bibinfo  {journal} {Nature materials}\ }\textbf {\bibinfo
  {volume} {11}},\ \bibinfo {pages} {565} (\bibinfo {year} {2012})}\BibitemShut
  {NoStop}%
\bibitem [{\citenamefont {Wegener}(2013)}]{Wegener2013}%
  \BibitemOpen
  \bibfield  {author} {\bibinfo {author} {\bibfnamefont {M.}~\bibnamefont
  {Wegener}},\ }\bibfield  {title} {\bibinfo {title} {{Metamaterials Beyond
  Optics}},\ }\href {https://doi.org/10.1126/SCIENCE.1246545} {\bibfield
  {journal} {\bibinfo  {journal} {Science}\ }\textbf {\bibinfo {volume}
  {342}},\ \bibinfo {pages} {939} (\bibinfo {year} {2013})}\BibitemShut
  {NoStop}%
\bibitem [{\citenamefont {Fernandez-Corbaton}\ \emph
  {et~al.}(2019)\citenamefont {Fernandez-Corbaton}, \citenamefont {Rockstuhl},
  \citenamefont {Ziemke}, \citenamefont {Gumbsch}, \citenamefont {Albiez},
  \citenamefont {Schwaiger}, \citenamefont {Frenzel}, \citenamefont {Kadic},\
  and\ \citenamefont {Wegener}}]{Fernandez-Corbaton2019}%
  \BibitemOpen
  \bibfield  {author} {\bibinfo {author} {\bibfnamefont {I.}~\bibnamefont
  {Fernandez-Corbaton}}, \bibinfo {author} {\bibfnamefont {C.}~\bibnamefont
  {Rockstuhl}}, \bibinfo {author} {\bibfnamefont {P.}~\bibnamefont {Ziemke}},
  \bibinfo {author} {\bibfnamefont {P.}~\bibnamefont {Gumbsch}}, \bibinfo
  {author} {\bibfnamefont {A.}~\bibnamefont {Albiez}}, \bibinfo {author}
  {\bibfnamefont {R.}~\bibnamefont {Schwaiger}}, \bibinfo {author}
  {\bibfnamefont {T.}~\bibnamefont {Frenzel}}, \bibinfo {author} {\bibfnamefont
  {M.}~\bibnamefont {Kadic}},\ and\ \bibinfo {author} {\bibfnamefont
  {M.}~\bibnamefont {Wegener}},\ }\bibfield  {title} {\bibinfo {title} {{New
  Twists of 3D Chiral Metamaterials}},\ }\href
  {https://doi.org/10.1002/ADMA.201807742} {\bibfield  {journal} {\bibinfo
  {journal} {Advanced Materials}\ }\textbf {\bibinfo {volume} {31}},\ \bibinfo
  {pages} {1807742} (\bibinfo {year} {2019})}\BibitemShut {NoStop}%
\bibitem [{\citenamefont {{Scheibner, Colin and Souslov, Anton and Banerjee,
  Debarghya and Sur{\'o}wka, Piotr and Irvine, William and Vitelli,
  Vincenzo}}(2020)}]{scheibner2020odd}%
  \BibitemOpen
  \bibfield  {author} {\bibinfo {author} {\bibnamefont {{Scheibner, Colin and
  Souslov, Anton and Banerjee, Debarghya and Sur{\'o}wka, Piotr and Irvine,
  William and Vitelli, Vincenzo}}},\ }\bibfield  {title} {\bibinfo {title}
  {{Odd elasticity}},\ }\href {https://doi.org/10.1038/s41567-020-0795-y}
  {\bibfield  {journal} {\bibinfo  {journal} {Nature Physics}\ }\textbf
  {\bibinfo {volume} {16}},\ \bibinfo {pages} {475} (\bibinfo {year}
  {2020})}\BibitemShut {NoStop}%
\bibitem [{\citenamefont {{Bertoldi, Katia and Vitelli, Vincenzo and
  Christensen, Johan and Van Hecke, Martin}}(2017)}]{bertoldi2017flexible}%
  \BibitemOpen
  \bibfield  {author} {\bibinfo {author} {\bibnamefont {{Bertoldi, Katia and
  Vitelli, Vincenzo and Christensen, Johan and Van Hecke, Martin}}},\
  }\bibfield  {title} {\bibinfo {title} {{Flexible mechanical metamaterials}},\
  }\href {https://doi.org/10.1038/natrevmats.2017.66} {\bibfield  {journal}
  {\bibinfo  {journal} {Nature Reviews Materials}\ }\textbf {\bibinfo {volume}
  {2}},\ \bibinfo {pages} {1} (\bibinfo {year} {2017})}\BibitemShut {NoStop}%
\bibitem [{\citenamefont {Fischer}\ \emph {et~al.}(2020)\citenamefont
  {Fischer}, \citenamefont {Hillen},\ and\ \citenamefont
  {Eberl}}]{Fischer2020}%
  \BibitemOpen
  \bibfield  {author} {\bibinfo {author} {\bibfnamefont {S.~C.}\ \bibnamefont
  {Fischer}}, \bibinfo {author} {\bibfnamefont {L.}~\bibnamefont {Hillen}},\
  and\ \bibinfo {author} {\bibfnamefont {C.}~\bibnamefont {Eberl}},\ }\bibfield
   {title} {\bibinfo {title} {{Mechanical Metamaterials on the Way from
  Laboratory Scale to Industrial Applications: Challenges for Characterization
  and Scalability}},\ }\href {https://doi.org/10.3390/MA13163605} {\bibfield
  {journal} {\bibinfo  {journal} {Materials 2020, Vol. 13, Page 3605}\ }\textbf
  {\bibinfo {volume} {13}},\ \bibinfo {pages} {3605} (\bibinfo {year}
  {2020})}\BibitemShut {NoStop}%
\bibitem [{\citenamefont {{Neville, Robin M and Scarpa, Fabrizio and Pirrera,
  Alberto}}(2016)}]{neville2016shape}%
  \BibitemOpen
  \bibfield  {author} {\bibinfo {author} {\bibnamefont {{Neville, Robin M and
  Scarpa, Fabrizio and Pirrera, Alberto}}},\ }\bibfield  {title} {\bibinfo
  {title} {{Shape morphing Kirigami mechanical metamaterials}},\ }\href
  {https://doi.org/10.1038/srep31067} {\bibfield  {journal} {\bibinfo
  {journal} {Scientific reports}\ }\textbf {\bibinfo {volume} {6}},\ \bibinfo
  {pages} {1} (\bibinfo {year} {2016})}\BibitemShut {NoStop}%
\bibitem [{\citenamefont {Florijn}\ \emph {et~al.}(2014)\citenamefont
  {Florijn}, \citenamefont {Coulais},\ and\ \citenamefont {{Van
  Hecke}}}]{Florijn2014}%
  \BibitemOpen
  \bibfield  {author} {\bibinfo {author} {\bibfnamefont {B.}~\bibnamefont
  {Florijn}}, \bibinfo {author} {\bibfnamefont {C.}~\bibnamefont {Coulais}},\
  and\ \bibinfo {author} {\bibfnamefont {M.}~\bibnamefont {{Van Hecke}}},\
  }\bibfield  {title} {\bibinfo {title} {{Programmable mechanical
  metamaterials}},\ }\href {https://doi.org/10.1103/PhysRevLett.113.175503}
  {\bibfield  {journal} {\bibinfo  {journal} {Physical Review Letters}\
  }\textbf {\bibinfo {volume} {113}},\ \bibinfo {pages} {175503} (\bibinfo
  {year} {2014})}\BibitemShut {NoStop}%
\bibitem [{\citenamefont {Schaedler}\ \emph {et~al.}(2011)\citenamefont
  {Schaedler}, \citenamefont {Jacobsen}, \citenamefont {Torrents},
  \citenamefont {Sorensen}, \citenamefont {Lian}, \citenamefont {Greer},
  \citenamefont {Valdevit},\ and\ \citenamefont {Carter}}]{Schaedler2011}%
  \BibitemOpen
  \bibfield  {author} {\bibinfo {author} {\bibfnamefont {T.~A.}\ \bibnamefont
  {Schaedler}}, \bibinfo {author} {\bibfnamefont {A.~J.}\ \bibnamefont
  {Jacobsen}}, \bibinfo {author} {\bibfnamefont {A.}~\bibnamefont {Torrents}},
  \bibinfo {author} {\bibfnamefont {A.~E.}\ \bibnamefont {Sorensen}}, \bibinfo
  {author} {\bibfnamefont {J.}~\bibnamefont {Lian}}, \bibinfo {author}
  {\bibfnamefont {J.~R.}\ \bibnamefont {Greer}}, \bibinfo {author}
  {\bibfnamefont {L.}~\bibnamefont {Valdevit}},\ and\ \bibinfo {author}
  {\bibfnamefont {W.~B.}\ \bibnamefont {Carter}},\ }\bibfield  {title}
  {\bibinfo {title} {{Ultralight metallic microlattices}},\ }\href
  {https://doi.org/10.1126/science.1211649} {\bibfield  {journal} {\bibinfo
  {journal} {Science}\ }\textbf {\bibinfo {volume} {334}},\ \bibinfo {pages}
  {962} (\bibinfo {year} {2011})}\BibitemShut {NoStop}%
\bibitem [{\citenamefont {Pendry}\ \emph {et~al.}(2006)\citenamefont {Pendry},
  \citenamefont {Schurig},\ and\ \citenamefont {Smith}}]{Pendry2006}%
  \BibitemOpen
  \bibfield  {author} {\bibinfo {author} {\bibfnamefont {J.~B.}\ \bibnamefont
  {Pendry}}, \bibinfo {author} {\bibfnamefont {D.}~\bibnamefont {Schurig}},\
  and\ \bibinfo {author} {\bibfnamefont {D.~R.}\ \bibnamefont {Smith}},\
  }\bibfield  {title} {\bibinfo {title} {{Controlling electromagnetic
  fields}},\ }\href {https://doi.org/10.1126/science.1125907} {\bibfield
  {journal} {\bibinfo  {journal} {Science}\ }\textbf {\bibinfo {volume}
  {312}},\ \bibinfo {pages} {1780} (\bibinfo {year} {2006})}\BibitemShut
  {NoStop}%
\bibitem [{\citenamefont {B{\"{u}}ckmann}\ \emph {et~al.}(2014)\citenamefont
  {B{\"{u}}ckmann}, \citenamefont {Thiel}, \citenamefont {Kadic}, \citenamefont
  {Schittny},\ and\ \citenamefont {Wegener}}]{Buckmann2014}%
  \BibitemOpen
  \bibfield  {author} {\bibinfo {author} {\bibfnamefont {T.}~\bibnamefont
  {B{\"{u}}ckmann}}, \bibinfo {author} {\bibfnamefont {M.}~\bibnamefont
  {Thiel}}, \bibinfo {author} {\bibfnamefont {M.}~\bibnamefont {Kadic}},
  \bibinfo {author} {\bibfnamefont {R.}~\bibnamefont {Schittny}},\ and\
  \bibinfo {author} {\bibfnamefont {M.}~\bibnamefont {Wegener}},\ }\bibfield
  {title} {\bibinfo {title} {{An elasto-mechanical unfeelability cloak made of
  pentamode metamaterials}},\ }\href {https://doi.org/10.1038/ncomms5130}
  {\bibfield  {journal} {\bibinfo  {journal} {Nature Communications 2014 5:1}\
  }\textbf {\bibinfo {volume} {5}},\ \bibinfo {pages} {1} (\bibinfo {year}
  {2014})}\BibitemShut {NoStop}%
\bibitem [{\citenamefont {Lakes}(1987)}]{Lakes1987}%
  \BibitemOpen
  \bibfield  {author} {\bibinfo {author} {\bibfnamefont {R.}~\bibnamefont
  {Lakes}},\ }\bibfield  {title} {\bibinfo {title} {{Foam Structures with a
  Negative Poisson's Ratio}},\ }\href
  {https://doi.org/10.1126/SCIENCE.235.4792.1038} {\bibfield  {journal}
  {\bibinfo  {journal} {Science}\ }\textbf {\bibinfo {volume} {235}},\ \bibinfo
  {pages} {1038} (\bibinfo {year} {1987})}\BibitemShut {NoStop}%
\bibitem [{\citenamefont {Alderson}\ and\ \citenamefont
  {Evans}(2001)}]{Alderson2001}%
  \BibitemOpen
  \bibfield  {author} {\bibinfo {author} {\bibfnamefont {A.}~\bibnamefont
  {Alderson}}\ and\ \bibinfo {author} {\bibfnamefont {K.~E.}\ \bibnamefont
  {Evans}},\ }\bibfield  {title} {\bibinfo {title} {{Rotation and dilation
  deformation mechanisms for auxetic behaviour in the $\alpha$-cristobalite
  tetrahedral framework structure}},\ }\href
  {https://doi.org/10.1007/S002690100209} {\bibfield  {journal} {\bibinfo
  {journal} {Physics and Chemistry of Minerals 2001 28:10}\ }\textbf {\bibinfo
  {volume} {28}},\ \bibinfo {pages} {711} (\bibinfo {year} {2001})}\BibitemShut
  {NoStop}%
\bibitem [{\citenamefont {Mizzi}\ \emph {et~al.}(2018)\citenamefont {Mizzi},
  \citenamefont {Mahdi}, \citenamefont {Titov}, \citenamefont {Gatt},
  \citenamefont {Attard}, \citenamefont {Evans}, \citenamefont {Grima},\ and\
  \citenamefont {Tan}}]{Mizzi2018}%
  \BibitemOpen
  \bibfield  {author} {\bibinfo {author} {\bibfnamefont {L.}~\bibnamefont
  {Mizzi}}, \bibinfo {author} {\bibfnamefont {E.~M.}\ \bibnamefont {Mahdi}},
  \bibinfo {author} {\bibfnamefont {K.}~\bibnamefont {Titov}}, \bibinfo
  {author} {\bibfnamefont {R.}~\bibnamefont {Gatt}}, \bibinfo {author}
  {\bibfnamefont {D.}~\bibnamefont {Attard}}, \bibinfo {author} {\bibfnamefont
  {K.~E.}\ \bibnamefont {Evans}}, \bibinfo {author} {\bibfnamefont {J.~N.}\
  \bibnamefont {Grima}},\ and\ \bibinfo {author} {\bibfnamefont {J.~C.}\
  \bibnamefont {Tan}},\ }\bibfield  {title} {\bibinfo {title} {{Mechanical
  metamaterials with star-shaped pores exhibiting negative and zero Poisson's
  ratio}},\ }\href {https://doi.org/10.1016/J.MATDES.2018.02.051} {\bibfield
  {journal} {\bibinfo  {journal} {Materials \& Design}\ }\textbf {\bibinfo
  {volume} {146}},\ \bibinfo {pages} {28} (\bibinfo {year} {2018})}\BibitemShut
  {NoStop}%
\bibitem [{\citenamefont {Duan}\ \emph {et~al.}(2020)\citenamefont {Duan},
  \citenamefont {Xi}, \citenamefont {Wen},\ and\ \citenamefont
  {Fang}}]{Duan2020}%
  \BibitemOpen
  \bibfield  {author} {\bibinfo {author} {\bibfnamefont {S.}~\bibnamefont
  {Duan}}, \bibinfo {author} {\bibfnamefont {L.}~\bibnamefont {Xi}}, \bibinfo
  {author} {\bibfnamefont {W.}~\bibnamefont {Wen}},\ and\ \bibinfo {author}
  {\bibfnamefont {D.}~\bibnamefont {Fang}},\ }\bibfield  {title} {\bibinfo
  {title} {{A novel design method for 3D positive and negative Poisson's ratio
  material based on tension-twist coupling effects}},\ }\href
  {https://doi.org/10.1016/J.COMPSTRUCT.2020.111899} {\bibfield  {journal}
  {\bibinfo  {journal} {Composite Structures}\ }\textbf {\bibinfo {volume}
  {236}},\ \bibinfo {pages} {111899} (\bibinfo {year} {2020})}\BibitemShut
  {NoStop}%
\bibitem [{\citenamefont {Baughman}\ \emph {et~al.}(1998)\citenamefont
  {Baughman}, \citenamefont {Stafstr{\"{o}}m}, \citenamefont {Cui},\ and\
  \citenamefont {Dantas}}]{Baughman1998}%
  \BibitemOpen
  \bibfield  {author} {\bibinfo {author} {\bibfnamefont {R.~H.}\ \bibnamefont
  {Baughman}}, \bibinfo {author} {\bibfnamefont {S.}~\bibnamefont
  {Stafstr{\"{o}}m}}, \bibinfo {author} {\bibfnamefont {C.}~\bibnamefont
  {Cui}},\ and\ \bibinfo {author} {\bibfnamefont {S.~O.}\ \bibnamefont
  {Dantas}},\ }\bibfield  {title} {\bibinfo {title} {{Materials with Negative
  Compressibilities in One or More Dimensions}},\ }\href
  {https://doi.org/10.1126/SCIENCE.279.5356.1522} {\bibfield  {journal}
  {\bibinfo  {journal} {Science}\ }\textbf {\bibinfo {volume} {279}},\ \bibinfo
  {pages} {1522} (\bibinfo {year} {1998})}\BibitemShut {NoStop}%
\bibitem [{\citenamefont {Lakes}\ and\ \citenamefont
  {Wojciechowski}(2008)}]{lakes2008negative}%
  \BibitemOpen
  \bibfield  {author} {\bibinfo {author} {\bibfnamefont {R.}~\bibnamefont
  {Lakes}}\ and\ \bibinfo {author} {\bibfnamefont {K.}~\bibnamefont
  {Wojciechowski}},\ }\bibfield  {title} {\bibinfo {title} {Negative
  compressibility, negative poisson's ratio, and stability},\ }\href
  {https://doi.org/10.1002/pssb.200777708} {\bibfield  {journal} {\bibinfo
  {journal} {physica status solidi (b)}\ }\textbf {\bibinfo {volume} {245}},\
  \bibinfo {pages} {545} (\bibinfo {year} {2008})}\BibitemShut {NoStop}%
\bibitem [{\citenamefont {Janmey}\ \emph {et~al.}(2006)\citenamefont {Janmey},
  \citenamefont {McCormick}, \citenamefont {Rammensee}, \citenamefont {Leight},
  \citenamefont {Georges},\ and\ \citenamefont {MacKintosh}}]{Janmey2006}%
  \BibitemOpen
  \bibfield  {author} {\bibinfo {author} {\bibfnamefont {P.~A.}\ \bibnamefont
  {Janmey}}, \bibinfo {author} {\bibfnamefont {M.~E.}\ \bibnamefont
  {McCormick}}, \bibinfo {author} {\bibfnamefont {S.}~\bibnamefont
  {Rammensee}}, \bibinfo {author} {\bibfnamefont {J.~L.}\ \bibnamefont
  {Leight}}, \bibinfo {author} {\bibfnamefont {P.~C.}\ \bibnamefont
  {Georges}},\ and\ \bibinfo {author} {\bibfnamefont {F.~C.}\ \bibnamefont
  {MacKintosh}},\ }\bibfield  {title} {\bibinfo {title} {{Negative normal
  stress in semiflexible biopolymer gels}},\ }\href
  {https://doi.org/10.1038/nmat1810} {\bibfield  {journal} {\bibinfo  {journal}
  {Nature Materials 2007 6:1}\ }\textbf {\bibinfo {volume} {6}},\ \bibinfo
  {pages} {48} (\bibinfo {year} {2006})}\BibitemShut {NoStop}%
\bibitem [{\citenamefont {Yuan}\ \emph {et~al.}(2021)\citenamefont {Yuan},
  \citenamefont {Chen}, \citenamefont {Yao}, \citenamefont {Guo}, \citenamefont
  {Huang}, \citenamefont {Peng}, \citenamefont {Xu}, \citenamefont {Lv},
  \citenamefont {Tao}, \citenamefont {Duan} \emph {et~al.}}]{Yuan2021}%
  \BibitemOpen
  \bibfield  {author} {\bibinfo {author} {\bibfnamefont {X.}~\bibnamefont
  {Yuan}}, \bibinfo {author} {\bibfnamefont {M.}~\bibnamefont {Chen}}, \bibinfo
  {author} {\bibfnamefont {Y.}~\bibnamefont {Yao}}, \bibinfo {author}
  {\bibfnamefont {X.}~\bibnamefont {Guo}}, \bibinfo {author} {\bibfnamefont
  {Y.}~\bibnamefont {Huang}}, \bibinfo {author} {\bibfnamefont
  {Z.}~\bibnamefont {Peng}}, \bibinfo {author} {\bibfnamefont {B.}~\bibnamefont
  {Xu}}, \bibinfo {author} {\bibfnamefont {B.}~\bibnamefont {Lv}}, \bibinfo
  {author} {\bibfnamefont {R.}~\bibnamefont {Tao}}, \bibinfo {author}
  {\bibfnamefont {S.}~\bibnamefont {Duan}}, \emph {et~al.},\ }\bibfield
  {title} {\bibinfo {title} {{Recent progress in the design and fabrication of
  multifunctional structures based on metamaterials}},\ }\href
  {https://doi.org/10.1016/J.COSSMS.2020.100883} {\bibfield  {journal}
  {\bibinfo  {journal} {Current Opinion in Solid State and Materials Science}\
  }\textbf {\bibinfo {volume} {25}},\ \bibinfo {pages} {100883} (\bibinfo
  {year} {2021})}\BibitemShut {NoStop}%
\bibitem [{\citenamefont {Ha}\ \emph {et~al.}(2016)\citenamefont {Ha},
  \citenamefont {Plesha},\ and\ \citenamefont {Lakes}}]{ha2016chiral}%
  \BibitemOpen
  \bibfield  {author} {\bibinfo {author} {\bibfnamefont {C.~S.}\ \bibnamefont
  {Ha}}, \bibinfo {author} {\bibfnamefont {M.~E.}\ \bibnamefont {Plesha}},\
  and\ \bibinfo {author} {\bibfnamefont {R.~S.}\ \bibnamefont {Lakes}},\
  }\bibfield  {title} {\bibinfo {title} {Chiral three-dimensional isotropic
  lattices with negative poisson's ratio},\ }\href
  {https://doi.org/10.1002/pssb.201600055} {\bibfield  {journal} {\bibinfo
  {journal} {physica status solidi (b)}\ }\textbf {\bibinfo {volume} {253}},\
  \bibinfo {pages} {1243} (\bibinfo {year} {2016})}\BibitemShut {NoStop}%
\bibitem [{\citenamefont {Frenzel}\ \emph {et~al.}(2017)\citenamefont
  {Frenzel}, \citenamefont {Kadic},\ and\ \citenamefont
  {Wegener}}]{Frenzel2017}%
  \BibitemOpen
  \bibfield  {author} {\bibinfo {author} {\bibfnamefont {T.}~\bibnamefont
  {Frenzel}}, \bibinfo {author} {\bibfnamefont {M.}~\bibnamefont {Kadic}},\
  and\ \bibinfo {author} {\bibfnamefont {M.}~\bibnamefont {Wegener}},\
  }\bibfield  {title} {\bibinfo {title} {{Three-dimensional mechanical
  metamaterials with a twist}},\ }\href
  {https://doi.org/10.1126/science.aao4640} {\bibfield  {journal} {\bibinfo
  {journal} {Science}\ }\textbf {\bibinfo {volume} {358}},\ \bibinfo {pages}
  {1072} (\bibinfo {year} {2017})}\BibitemShut {NoStop}%
\bibitem [{\citenamefont {Chen}\ \emph {et~al.}(2018)\citenamefont {Chen},
  \citenamefont {Ruan},\ and\ \citenamefont {Huang}}]{Chen2018}%
  \BibitemOpen
  \bibfield  {author} {\bibinfo {author} {\bibfnamefont {W.}~\bibnamefont
  {Chen}}, \bibinfo {author} {\bibfnamefont {D.}~\bibnamefont {Ruan}},\ and\
  \bibinfo {author} {\bibfnamefont {X.}~\bibnamefont {Huang}},\ }\bibfield
  {title} {\bibinfo {title} {{Optimization for twist chirality of structural
  materials induced by axial strain}},\ }\href
  {https://doi.org/10.1016/J.MTCOMM.2018.03.010} {\bibfield  {journal}
  {\bibinfo  {journal} {Materials Today Communications}\ }\textbf {\bibinfo
  {volume} {15}},\ \bibinfo {pages} {175} (\bibinfo {year} {2018})}\BibitemShut
  {NoStop}%
\bibitem [{\citenamefont {Wu}\ \emph {et~al.}(2018)\citenamefont {Wu},
  \citenamefont {Geng}, \citenamefont {Niu}, \citenamefont {Qi}, \citenamefont
  {Cui},\ and\ \citenamefont {Fang}}]{Wu2018}%
  \BibitemOpen
  \bibfield  {author} {\bibinfo {author} {\bibfnamefont {W.}~\bibnamefont
  {Wu}}, \bibinfo {author} {\bibfnamefont {L.}~\bibnamefont {Geng}}, \bibinfo
  {author} {\bibfnamefont {Y.}~\bibnamefont {Niu}}, \bibinfo {author}
  {\bibfnamefont {D.}~\bibnamefont {Qi}}, \bibinfo {author} {\bibfnamefont
  {X.}~\bibnamefont {Cui}},\ and\ \bibinfo {author} {\bibfnamefont
  {D.}~\bibnamefont {Fang}},\ }\bibfield  {title} {\bibinfo {title}
  {{Compression twist deformation of novel tetrachiral architected cylindrical
  tube inspired by towel gourd tendrils}},\ }\href
  {https://doi.org/10.1016/J.EML.2018.02.001} {\bibfield  {journal} {\bibinfo
  {journal} {Extreme Mechanics Letters}\ }\textbf {\bibinfo {volume} {20}},\
  \bibinfo {pages} {104} (\bibinfo {year} {2018})}\BibitemShut {NoStop}%
\bibitem [{\citenamefont {Ma}\ \emph {et~al.}(2018)\citenamefont {Ma},
  \citenamefont {Lei}, \citenamefont {Hua}, \citenamefont {Bai}, \citenamefont
  {Liang},\ and\ \citenamefont {Fang}}]{Ma2018}%
  \BibitemOpen
  \bibfield  {author} {\bibinfo {author} {\bibfnamefont {C.}~\bibnamefont
  {Ma}}, \bibinfo {author} {\bibfnamefont {H.}~\bibnamefont {Lei}}, \bibinfo
  {author} {\bibfnamefont {J.}~\bibnamefont {Hua}}, \bibinfo {author}
  {\bibfnamefont {Y.}~\bibnamefont {Bai}}, \bibinfo {author} {\bibfnamefont
  {J.}~\bibnamefont {Liang}},\ and\ \bibinfo {author} {\bibfnamefont
  {D.}~\bibnamefont {Fang}},\ }\bibfield  {title} {\bibinfo {title}
  {{Experimental and simulation investigation of the reversible bi-directional
  twisting response of tetra-chiral cylindrical shells}},\ }\href
  {https://doi.org/10.1016/J.COMPSTRUCT.2018.07.013} {\bibfield  {journal}
  {\bibinfo  {journal} {Composite Structures}\ }\textbf {\bibinfo {volume}
  {203}},\ \bibinfo {pages} {142} (\bibinfo {year} {2018})}\BibitemShut
  {NoStop}%
\bibitem [{\citenamefont {Fu}\ \emph {et~al.}(2018)\citenamefont {Fu},
  \citenamefont {Liu},\ and\ \citenamefont {Hu}}]{Fu2018}%
  \BibitemOpen
  \bibfield  {author} {\bibinfo {author} {\bibfnamefont {M.}~\bibnamefont
  {Fu}}, \bibinfo {author} {\bibfnamefont {F.}~\bibnamefont {Liu}},\ and\
  \bibinfo {author} {\bibfnamefont {L.}~\bibnamefont {Hu}},\ }\bibfield
  {title} {\bibinfo {title} {{A novel category of 3D chiral material with
  negative Poisson's ratio}},\ }\href
  {https://doi.org/10.1016/J.COMPSCITECH.2018.03.017} {\bibfield  {journal}
  {\bibinfo  {journal} {Composites Science and Technology}\ }\textbf {\bibinfo
  {volume} {160}},\ \bibinfo {pages} {111} (\bibinfo {year}
  {2018})}\BibitemShut {NoStop}%
\bibitem [{\citenamefont {Zheng}\ \emph {et~al.}(2019)\citenamefont {Zheng},
  \citenamefont {Zhong}, \citenamefont {Chen}, \citenamefont {Fu},\ and\
  \citenamefont {Hu}}]{Zheng2019}%
  \BibitemOpen
  \bibfield  {author} {\bibinfo {author} {\bibfnamefont {B.~B.}\ \bibnamefont
  {Zheng}}, \bibinfo {author} {\bibfnamefont {R.~C.}\ \bibnamefont {Zhong}},
  \bibinfo {author} {\bibfnamefont {X.}~\bibnamefont {Chen}}, \bibinfo {author}
  {\bibfnamefont {M.~H.}\ \bibnamefont {Fu}},\ and\ \bibinfo {author}
  {\bibfnamefont {L.~L.}\ \bibnamefont {Hu}},\ }\bibfield  {title} {\bibinfo
  {title} {{A novel metamaterial with tension-torsion coupling effect}},\
  }\href {https://doi.org/10.1016/J.MATDES.2019.107700} {\bibfield  {journal}
  {\bibinfo  {journal} {Materials \& Design}\ }\textbf {\bibinfo {volume}
  {171}},\ \bibinfo {pages} {107700} (\bibinfo {year} {2019})}\BibitemShut
  {NoStop}%
\bibitem [{\citenamefont {Zhong}\ \emph {et~al.}(2019)\citenamefont {Zhong},
  \citenamefont {Fu}, \citenamefont {Chen}, \citenamefont {Zheng},\ and\
  \citenamefont {Hu}}]{Zhong2019}%
  \BibitemOpen
  \bibfield  {author} {\bibinfo {author} {\bibfnamefont {R.}~\bibnamefont
  {Zhong}}, \bibinfo {author} {\bibfnamefont {M.}~\bibnamefont {Fu}}, \bibinfo
  {author} {\bibfnamefont {X.}~\bibnamefont {Chen}}, \bibinfo {author}
  {\bibfnamefont {B.}~\bibnamefont {Zheng}},\ and\ \bibinfo {author}
  {\bibfnamefont {L.}~\bibnamefont {Hu}},\ }\bibfield  {title} {\bibinfo
  {title} {{A novel three-dimensional mechanical metamaterial with
  compression-torsion properties}},\ }\href
  {https://doi.org/10.1016/J.COMPSTRUCT.2019.111232} {\bibfield  {journal}
  {\bibinfo  {journal} {Composite Structures}\ }\textbf {\bibinfo {volume}
  {226}},\ \bibinfo {pages} {111232} (\bibinfo {year} {2019})}\BibitemShut
  {NoStop}%
\bibitem [{\citenamefont {Li}\ \emph {et~al.}(2019)\citenamefont {Li},
  \citenamefont {Yang},\ and\ \citenamefont {Lu}}]{Li2019}%
  \BibitemOpen
  \bibfield  {author} {\bibinfo {author} {\bibfnamefont {X.}~\bibnamefont
  {Li}}, \bibinfo {author} {\bibfnamefont {Z.}~\bibnamefont {Yang}},\ and\
  \bibinfo {author} {\bibfnamefont {Z.}~\bibnamefont {Lu}},\ }\bibfield
  {title} {\bibinfo {title} {{Design 3D metamaterials with
  compression-induced-twisting characteristics using shear–compression
  coupling effects}},\ }\href {https://doi.org/10.1016/J.EML.2019.100471}
  {\bibfield  {journal} {\bibinfo  {journal} {Extreme Mechanics Letters}\
  }\textbf {\bibinfo {volume} {29}},\ \bibinfo {pages} {100471} (\bibinfo
  {year} {2019})}\BibitemShut {NoStop}%
\bibitem [{\citenamefont {Wang}\ \emph {et~al.}(2020)\citenamefont {Wang},
  \citenamefont {Liu},\ and\ \citenamefont {Zhang}}]{Wang2020}%
  \BibitemOpen
  \bibfield  {author} {\bibinfo {author} {\bibfnamefont {Y.~B.}\ \bibnamefont
  {Wang}}, \bibinfo {author} {\bibfnamefont {H.~T.}\ \bibnamefont {Liu}},\ and\
  \bibinfo {author} {\bibfnamefont {Z.~Y.}\ \bibnamefont {Zhang}},\ }\bibfield
  {title} {\bibinfo {title} {{Rotation spring: Rotation symmetric
  compression-torsion conversion structure with high space utilization}},\
  }\href {https://doi.org/10.1016/J.COMPSTRUCT.2020.112341} {\bibfield
  {journal} {\bibinfo  {journal} {Composite Structures}\ }\textbf {\bibinfo
  {volume} {245}},\ \bibinfo {pages} {112341} (\bibinfo {year}
  {2020})}\BibitemShut {NoStop}%
\bibitem [{\citenamefont {Wang}\ and\ \citenamefont {Liu}(2020)}]{Wang2020a}%
  \BibitemOpen
  \bibfield  {author} {\bibinfo {author} {\bibfnamefont {L.}~\bibnamefont
  {Wang}}\ and\ \bibinfo {author} {\bibfnamefont {H.~T.}\ \bibnamefont {Liu}},\
  }\bibfield  {title} {\bibinfo {title} {{3D compression–torsion cubic
  mechanical metamaterial with double inclined rods}},\ }\href
  {https://doi.org/10.1016/J.EML.2020.100706} {\bibfield  {journal} {\bibinfo
  {journal} {Extreme Mechanics Letters}\ }\textbf {\bibinfo {volume} {37}},\
  \bibinfo {pages} {100706} (\bibinfo {year} {2020})}\BibitemShut {NoStop}%
\bibitem [{\citenamefont {Frenzel}\ \emph {et~al.}(2021)\citenamefont
  {Frenzel}, \citenamefont {Hahn}, \citenamefont {Ziemke}, \citenamefont
  {Schneider}, \citenamefont {Chen}, \citenamefont {Kiefer}, \citenamefont
  {Gumbsch},\ and\ \citenamefont {Wegener}}]{Frenzel2021}%
  \BibitemOpen
  \bibfield  {author} {\bibinfo {author} {\bibfnamefont {T.}~\bibnamefont
  {Frenzel}}, \bibinfo {author} {\bibfnamefont {V.}~\bibnamefont {Hahn}},
  \bibinfo {author} {\bibfnamefont {P.}~\bibnamefont {Ziemke}}, \bibinfo
  {author} {\bibfnamefont {J.~L.~G.}\ \bibnamefont {Schneider}}, \bibinfo
  {author} {\bibfnamefont {Y.}~\bibnamefont {Chen}}, \bibinfo {author}
  {\bibfnamefont {P.}~\bibnamefont {Kiefer}}, \bibinfo {author} {\bibfnamefont
  {P.}~\bibnamefont {Gumbsch}},\ and\ \bibinfo {author} {\bibfnamefont
  {M.}~\bibnamefont {Wegener}},\ }\bibfield  {title} {\bibinfo {title} {{Large
  characteristic lengths in 3D chiral elastic metamaterials}},\ }\href
  {https://doi.org/10.1038/s43246-020-00107-w} {\bibfield  {journal} {\bibinfo
  {journal} {Communications Materials 2021 2:1}\ }\textbf {\bibinfo {volume}
  {2}},\ \bibinfo {pages} {1} (\bibinfo {year} {2021})}\BibitemShut {NoStop}%
\bibitem [{\citenamefont {Wang}\ \emph {et~al.}(2021)\citenamefont {Wang},
  \citenamefont {Liu},\ and\ \citenamefont {Zhang}}]{Wang2021}%
  \BibitemOpen
  \bibfield  {author} {\bibinfo {author} {\bibfnamefont {Y.~B.}\ \bibnamefont
  {Wang}}, \bibinfo {author} {\bibfnamefont {H.~T.}\ \bibnamefont {Liu}},\ and\
  \bibinfo {author} {\bibfnamefont {D.~Q.}\ \bibnamefont {Zhang}},\ }\bibfield
  {title} {\bibinfo {title} {{Compression-torsion conversion behavior of a
  cylindrical mechanical metamaterial based on askew re-entrant cells}},\
  }\href {https://doi.org/10.1016/J.MATLET.2021.130572} {\bibfield  {journal}
  {\bibinfo  {journal} {Materials Letters}\ }\textbf {\bibinfo {volume}
  {303}},\ \bibinfo {pages} {130572} (\bibinfo {year} {2021})}\BibitemShut
  {NoStop}%
\bibitem [{\citenamefont {Surjadi}\ \emph {et~al.}(2019)\citenamefont
  {Surjadi}, \citenamefont {Gao}, \citenamefont {Du}, \citenamefont {Li},
  \citenamefont {Xiong}, \citenamefont {Fang},\ and\ \citenamefont
  {Lu}}]{Surjadi2019}%
  \BibitemOpen
  \bibfield  {author} {\bibinfo {author} {\bibfnamefont {J.~U.}\ \bibnamefont
  {Surjadi}}, \bibinfo {author} {\bibfnamefont {L.}~\bibnamefont {Gao}},
  \bibinfo {author} {\bibfnamefont {H.}~\bibnamefont {Du}}, \bibinfo {author}
  {\bibfnamefont {X.}~\bibnamefont {Li}}, \bibinfo {author} {\bibfnamefont
  {X.}~\bibnamefont {Xiong}}, \bibinfo {author} {\bibfnamefont {N.~X.}\
  \bibnamefont {Fang}},\ and\ \bibinfo {author} {\bibfnamefont
  {Y.}~\bibnamefont {Lu}},\ }\bibfield  {title} {\bibinfo {title} {{Mechanical
  Metamaterials and Their Engineering Applications}},\ }\href
  {https://doi.org/10.1002/ADEM.201800864} {\bibfield  {journal} {\bibinfo
  {journal} {Advanced Engineering Materials}\ }\textbf {\bibinfo {volume}
  {21}},\ \bibinfo {pages} {1800864} (\bibinfo {year} {2019})}\BibitemShut
  {NoStop}%
\bibitem [{\citenamefont {Mindlin}\ and\ \citenamefont
  {Eshel}(1968)}]{Mindlin1968}%
  \BibitemOpen
  \bibfield  {author} {\bibinfo {author} {\bibfnamefont {R.~D.}\ \bibnamefont
  {Mindlin}}\ and\ \bibinfo {author} {\bibfnamefont {N.~N.}\ \bibnamefont
  {Eshel}},\ }\bibfield  {title} {\bibinfo {title} {{On first strain-gradient
  theories in linear elasticity}},\ }\href
  {https://doi.org/10.1016/0020-7683(68)90036-X} {\bibfield  {journal}
  {\bibinfo  {journal} {International Journal of Solids and Structures}\
  }\textbf {\bibinfo {volume} {4}},\ \bibinfo {pages} {109} (\bibinfo {year}
  {1968})}\BibitemShut {NoStop}%
\bibitem [{\citenamefont {Lakes}\ and\ \citenamefont
  {Benedict}(1982)}]{lakes1982noncentrosymmetry}%
  \BibitemOpen
  \bibfield  {author} {\bibinfo {author} {\bibfnamefont {R.~S.}\ \bibnamefont
  {Lakes}}\ and\ \bibinfo {author} {\bibfnamefont {R.~L.}\ \bibnamefont
  {Benedict}},\ }\bibfield  {title} {\bibinfo {title} {Noncentrosymmetry in
  micropolar elasticity},\ }\href
  {https://doi.org/10.1016/0020-7225(82)90096-9} {\bibfield  {journal}
  {\bibinfo  {journal} {International Journal of Engineering Science}\ }\textbf
  {\bibinfo {volume} {20}},\ \bibinfo {pages} {1161} (\bibinfo {year}
  {1982})}\BibitemShut {NoStop}%
\bibitem [{\citenamefont {Eringen}(1999)}]{eringen1999theory}%
  \BibitemOpen
  \bibfield  {author} {\bibinfo {author} {\bibfnamefont {A.~C.}\ \bibnamefont
  {Eringen}},\ }\bibfield  {title} {\bibinfo {title} {Theory of micropolar
  elasticity},\ }in\ \href {https://doi.org/10.1007/978-1-4612-0555-5_5} {\emph
  {\bibinfo {booktitle} {Microcontinuum field theories}}}\ (\bibinfo
  {publisher} {Springer},\ \bibinfo {year} {1999})\ pp.\ \bibinfo {pages}
  {101--248}\BibitemShut {NoStop}%
\bibitem [{\citenamefont {Lakes}(1995)}]{lakes1995experimental}%
  \BibitemOpen
  \bibfield  {author} {\bibinfo {author} {\bibfnamefont {R.}~\bibnamefont
  {Lakes}},\ }\bibfield  {title} {\bibinfo {title} {Experimental methods for
  study of cosserat elastic solids and other generalized elastic continua},\
  }\href@noop {} {\bibfield  {journal} {\bibinfo  {journal} {Continuum models
  for materials with microstructure}\ }\textbf {\bibinfo {volume} {70}},\
  \bibinfo {pages} {1} (\bibinfo {year} {1995})}\BibitemShut {NoStop}%
\bibitem [{\citenamefont {Lakes}(2016)}]{lakes2016physical}%
  \BibitemOpen
  \bibfield  {author} {\bibinfo {author} {\bibfnamefont {R.}~\bibnamefont
  {Lakes}},\ }\bibfield  {title} {\bibinfo {title} {Physical meaning of elastic
  constants in cosserat, void, and microstretch elasticity},\ }\href
  {https://doi.org/10.2140/jomms.2016.11.217} {\bibfield  {journal} {\bibinfo
  {journal} {Journal of Mechanics of Materials and Structures}\ }\textbf
  {\bibinfo {volume} {11}},\ \bibinfo {pages} {217} (\bibinfo {year}
  {2016})}\BibitemShut {NoStop}%
\bibitem [{\citenamefont {Alavi}\ \emph {et~al.}(2021)\citenamefont {Alavi},
  \citenamefont {Nasimsobhan}, \citenamefont {Ganghoffer}, \citenamefont
  {Sinoimeri},\ and\ \citenamefont {Sadighi}}]{alavi2021chiral}%
  \BibitemOpen
  \bibfield  {author} {\bibinfo {author} {\bibfnamefont {S.}~\bibnamefont
  {Alavi}}, \bibinfo {author} {\bibfnamefont {M.}~\bibnamefont {Nasimsobhan}},
  \bibinfo {author} {\bibfnamefont {J.}~\bibnamefont {Ganghoffer}}, \bibinfo
  {author} {\bibfnamefont {A.}~\bibnamefont {Sinoimeri}},\ and\ \bibinfo
  {author} {\bibfnamefont {M.}~\bibnamefont {Sadighi}},\ }\bibfield  {title}
  {\bibinfo {title} {Chiral cosserat model for architected materials
  constructed by homogenization},\ }\href
  {https://doi.org/10.1007/s11012-021-01381-9} {\bibfield  {journal} {\bibinfo
  {journal} {Meccanica}\ }\textbf {\bibinfo {volume} {56}},\ \bibinfo {pages}
  {2547} (\bibinfo {year} {2021})}\BibitemShut {NoStop}%
\bibitem [{\citenamefont {Alavi}\ \emph
  {et~al.}(2022{\natexlab{a}})\citenamefont {Alavi}, \citenamefont
  {Ganghoffer}, \citenamefont {Sadighi}, \citenamefont {Nasimsobhan},\ and\
  \citenamefont {Akbarzadeh}}]{alavi2022continualization}%
  \BibitemOpen
  \bibfield  {author} {\bibinfo {author} {\bibfnamefont {S.}~\bibnamefont
  {Alavi}}, \bibinfo {author} {\bibfnamefont {J.}~\bibnamefont {Ganghoffer}},
  \bibinfo {author} {\bibfnamefont {M.}~\bibnamefont {Sadighi}}, \bibinfo
  {author} {\bibfnamefont {M.}~\bibnamefont {Nasimsobhan}},\ and\ \bibinfo
  {author} {\bibfnamefont {A.}~\bibnamefont {Akbarzadeh}},\ }\bibfield  {title}
  {\bibinfo {title} {Continualization method of lattice materials and analysis
  of size effects revisited based on cosserat models},\ }\href
  {https://doi.org/10.1016/j.ijsolstr.2022.111894} {\bibfield  {journal}
  {\bibinfo  {journal} {International Journal of Solids and Structures}\
  }\textbf {\bibinfo {volume} {254}},\ \bibinfo {pages} {111894} (\bibinfo
  {year} {2022}{\natexlab{a}})}\BibitemShut {NoStop}%
\bibitem [{\citenamefont {Duan}\ \emph {et~al.}(2018)\citenamefont {Duan},
  \citenamefont {Wen},\ and\ \citenamefont {Fang}}]{Duan2018}%
  \BibitemOpen
  \bibfield  {author} {\bibinfo {author} {\bibfnamefont {S.}~\bibnamefont
  {Duan}}, \bibinfo {author} {\bibfnamefont {W.}~\bibnamefont {Wen}},\ and\
  \bibinfo {author} {\bibfnamefont {D.}~\bibnamefont {Fang}},\ }\bibfield
  {title} {\bibinfo {title} {{A predictive micropolar continuum model for a
  novel three-dimensional chiral lattice with size effect and tension-twist
  coupling behavior}},\ }\href {https://doi.org/10.1016/J.JMPS.2018.07.016}
  {\bibfield  {journal} {\bibinfo  {journal} {Journal of the Mechanics and
  Physics of Solids}\ }\textbf {\bibinfo {volume} {121}},\ \bibinfo {pages}
  {23} (\bibinfo {year} {2018})}\BibitemShut {NoStop}%
\bibitem [{\citenamefont {Chen}\ \emph {et~al.}(2020)\citenamefont {Chen},
  \citenamefont {Frenzel}, \citenamefont {Guenneau}, \citenamefont {Kadic},\
  and\ \citenamefont {Wegener}}]{Chen2020}%
  \BibitemOpen
  \bibfield  {author} {\bibinfo {author} {\bibfnamefont {Y.}~\bibnamefont
  {Chen}}, \bibinfo {author} {\bibfnamefont {T.}~\bibnamefont {Frenzel}},
  \bibinfo {author} {\bibfnamefont {S.}~\bibnamefont {Guenneau}}, \bibinfo
  {author} {\bibfnamefont {M.}~\bibnamefont {Kadic}},\ and\ \bibinfo {author}
  {\bibfnamefont {M.}~\bibnamefont {Wegener}},\ }\bibfield  {title} {\bibinfo
  {title} {{Mapping acoustical activity in 3D chiral mechanical metamaterials
  onto micropolar continuum elasticity}},\ }\href
  {https://doi.org/10.1016/J.JMPS.2020.103877} {\bibfield  {journal} {\bibinfo
  {journal} {Journal of the Mechanics and Physics of Solids}\ }\textbf
  {\bibinfo {volume} {137}},\ \bibinfo {pages} {103877} (\bibinfo {year}
  {2020})}\BibitemShut {NoStop}%
\bibitem [{\citenamefont {Authier}(2013)}]{Authier2013}%
  \BibitemOpen
  \bibfield  {author} {\bibinfo {author} {\bibfnamefont {A.}~\bibnamefont
  {Authier}},\ }\href {https://doi.org/10.1107/97809553602060000900} {\emph
  {\bibinfo {title} {{Introduction to the properties of tensors}}}},\
  Vol.~\bibinfo {volume} {D}\ (\bibinfo  {publisher} {Wiley Online Library},\
  \bibinfo {year} {2013})\BibitemShut {NoStop}%
\bibitem [{\citenamefont {Milton}(2002)}]{Milton2002}%
  \BibitemOpen
  \bibfield  {author} {\bibinfo {author} {\bibfnamefont {G.~W.}\ \bibnamefont
  {Milton}},\ }\href {https://doi.org/10.1017/CBO9780511613357} {\emph
  {\bibinfo {title} {The Theory of Composites}}}\ (\bibinfo  {publisher}
  {Cambridge University Press},\ \bibinfo {year} {2002})\BibitemShut {NoStop}%
\bibitem [{\citenamefont {Ahamdi}\ and\ \citenamefont
  {Sohrabpour}(1999)}]{Ahamdi1999}%
  \BibitemOpen
  \bibfield  {author} {\bibinfo {author} {\bibfnamefont {G.}~\bibnamefont
  {Ahamdi}}\ and\ \bibinfo {author} {\bibfnamefont {S.}~\bibnamefont
  {Sohrabpour}},\ }\bibfield  {title} {\bibinfo {title} {{Theory of Micropolar
  Elasticity}},\ }\href {https://doi.org/10.1007/978-1-4612-0555-5_5}
  {\bibfield  {journal} {\bibinfo  {journal} {Iran J Sci Technol}\ }\textbf
  {\bibinfo {volume} {7}},\ \bibinfo {pages} {101} (\bibinfo {year}
  {1999})}\BibitemShut {NoStop}%
\bibitem [{\citenamefont {Goda}\ and\ \citenamefont
  {Ganghoffer}(2015)}]{Goda2015}%
  \BibitemOpen
  \bibfield  {author} {\bibinfo {author} {\bibfnamefont {I.}~\bibnamefont
  {Goda}}\ and\ \bibinfo {author} {\bibfnamefont {J.~F.}\ \bibnamefont
  {Ganghoffer}},\ }\bibfield  {title} {\bibinfo {title} {{Identification of
  couple-stress moduli of vertebral trabecular bone based on the 3D internal
  architectures}},\ }\href {https://doi.org/10.1016/J.JMBBM.2015.06.036}
  {\bibfield  {journal} {\bibinfo  {journal} {Journal of the Mechanical
  Behavior of Biomedical Materials}\ }\textbf {\bibinfo {volume} {51}},\
  \bibinfo {pages} {99} (\bibinfo {year} {2015})}\BibitemShut {NoStop}%
\bibitem [{\citenamefont {Karathanasopoulos}\ \emph {et~al.}(2017)\citenamefont
  {Karathanasopoulos}, \citenamefont {Reda},\ and\ \citenamefont
  {Ganghoffer}}]{karathanasopoulos2017designing}%
  \BibitemOpen
  \bibfield  {author} {\bibinfo {author} {\bibfnamefont {N.}~\bibnamefont
  {Karathanasopoulos}}, \bibinfo {author} {\bibfnamefont {H.}~\bibnamefont
  {Reda}},\ and\ \bibinfo {author} {\bibfnamefont {J.-f.}\ \bibnamefont
  {Ganghoffer}},\ }\bibfield  {title} {\bibinfo {title} {Designing
  two-dimensional metamaterials of controlled static and dynamic properties},\
  }\href {https://doi.org/10.1016/j.commatsci.2017.06.035} {\bibfield
  {journal} {\bibinfo  {journal} {Computational Materials Science}\ }\textbf
  {\bibinfo {volume} {138}},\ \bibinfo {pages} {323} (\bibinfo {year}
  {2017})}\BibitemShut {NoStop}%
\bibitem [{\citenamefont {Chen}\ and\ \citenamefont {Huang}(2019)}]{Chen2019}%
  \BibitemOpen
  \bibfield  {author} {\bibinfo {author} {\bibfnamefont {W.}~\bibnamefont
  {Chen}}\ and\ \bibinfo {author} {\bibfnamefont {X.}~\bibnamefont {Huang}},\
  }\bibfield  {title} {\bibinfo {title} {{Topological design of 3D chiral
  metamaterials based on couple-stress homogenization}},\ }\href
  {https://doi.org/10.1016/J.JMPS.2019.07.014} {\bibfield  {journal} {\bibinfo
  {journal} {Journal of the Mechanics and Physics of Solids}\ }\textbf
  {\bibinfo {volume} {131}},\ \bibinfo {pages} {372} (\bibinfo {year}
  {2019})}\BibitemShut {NoStop}%
\bibitem [{\citenamefont {Karathanasopoulos}\ \emph {et~al.}(2020)\citenamefont
  {Karathanasopoulos}, \citenamefont {Dos~Reis}, \citenamefont
  {Diamantopoulou},\ and\ \citenamefont
  {Ganghoffer}}]{karathanasopoulos2020mechanics}%
  \BibitemOpen
  \bibfield  {author} {\bibinfo {author} {\bibfnamefont {N.}~\bibnamefont
  {Karathanasopoulos}}, \bibinfo {author} {\bibfnamefont {F.}~\bibnamefont
  {Dos~Reis}}, \bibinfo {author} {\bibfnamefont {M.}~\bibnamefont
  {Diamantopoulou}},\ and\ \bibinfo {author} {\bibfnamefont {J.-F.}\
  \bibnamefont {Ganghoffer}},\ }\bibfield  {title} {\bibinfo {title} {Mechanics
  of beams made from chiral metamaterials: Tuning deflections through
  normal-shear strain couplings},\ }\href
  {https://doi.org/10.1016/j.matdes.2020.108520} {\bibfield  {journal}
  {\bibinfo  {journal} {Materials \& Design}\ }\textbf {\bibinfo {volume}
  {189}},\ \bibinfo {pages} {108520} (\bibinfo {year} {2020})}\BibitemShut
  {NoStop}%
\bibitem [{\citenamefont {Alavi}\ \emph
  {et~al.}(2022{\natexlab{b}})\citenamefont {Alavi}, \citenamefont
  {Ganghoffer},\ and\ \citenamefont {Sadighi}}]{alavi2022chiral}%
  \BibitemOpen
  \bibfield  {author} {\bibinfo {author} {\bibfnamefont {S.~E.}\ \bibnamefont
  {Alavi}}, \bibinfo {author} {\bibfnamefont {J.-F.}\ \bibnamefont
  {Ganghoffer}},\ and\ \bibinfo {author} {\bibfnamefont {M.}~\bibnamefont
  {Sadighi}},\ }\bibfield  {title} {\bibinfo {title} {Chiral cosserat
  homogenized constitutive models of architected media based on micromorphic
  homogenization},\ }\href {https://doi.org/10.1177/10812865221106941}
  {\bibfield  {journal} {\bibinfo  {journal} {Mathematics and Mechanics of
  Solids}\ }\textbf {\bibinfo {volume} {27}},\ \bibinfo {pages} {2287}
  (\bibinfo {year} {2022}{\natexlab{b}})}\BibitemShut {NoStop}%
\end{thebibliography}%

\end{document}